\documentclass[%
 reprint,
superscriptaddress,
nobibnotes,
 amsmath,amssymb,
 aps,
pre,
prstper,
floatfix,
]{revtex4-1}

\usepackage{graphicx}
\usepackage{epstopdf}
\usepackage{dcolumn}
\usepackage{bm}
\usepackage{hyperref}
\usepackage[mathlines]{lineno}
\usepackage{xcolor}
\usepackage{algorithmic}
\usepackage{amsmath}
\DeclareMathOperator{\sech}{sech}
\DeclareMathOperator{\csch}{csch}

\usepackage{array}

  \usepackage[caption=false,font=normalsize,labelfont=sf,textfont=sf]{subfig}

\usepackage{float}

\usepackage{graphicx}
\usepackage{grffile}
\usepackage{dcolumn}
\usepackage{multirow}
\usepackage{varioref}

\usepackage{appendix}

\begin{document}

\preprint{APS/123-QED}

\title{Intra-cavity field dynamics near avoided mode crossing in concentric silicon nitride ring resonator}

\author{Maitrayee Saha}
 \email{maitrayee@iitkgp.ac.in}
 \affiliation{
 Department of Physics, Indian Institute of Technology Kharagpur, Kharagpur-721302, India
 }
 \author{Samudra Roy}
 
 \affiliation{
 Department of Physics, Indian Institute of Technology Kharagpur, Kharagpur-721302, India
 }
 \affiliation{Centre of Theoretical Studies, Indian Institute of Technology Kharagpur, Kharagpur-721302, India}
 \author{ Shailendra K. Varshney}

 \renewcommand{\andname}{\ignorespaces}
 
\affiliation{
 Department of Electronics and Electrical Communication Engineering, Indian Institute of Technology Kharagpur, Kharagpur-721302, India
}

\date{\today}

\begin{abstract}
Understanding the intra-cavity field dynamics in passive microresonator systems has already been intriguing. It becomes fascinating when the system is complex, such as a concentric dual microring resonator that exhibit avoided mode crossing (AMC). In this work, we present a systematic study of intra-cavity oscillatory field dynamics near AMC in a concentric silicon nitride microring resonator with the help of coupled Lugiato-Lefever equation (LLE).  We identify two regions where the hybrid modes are strongly coupled and weakly coupled based on their eigenfrequency separation, which originate from mode coupling near AMC. We calculate the overlap integral to identify the cut-off pump wavelength region to observe the existence of the two hybrid modes. In strongly coupled hybrid mode region, we observe intra-cavity power oscillation and transfer of energy between the two hybrid modes in a periodic manner. We also evaluate the non-identical phases variation of these two modes. In weakly coupled hybrid mode region, power oscillation and energy transfer between modes reduces significantly , whereas their phase vary in almost identical fashion. We validate our numerical findings with the semi-analytical variational method, leading to an in-depth understanding of the mode coupling induced dynamics. We finally analyze the polarization properties of the field confined in the coupled system. Exploiting the Stokes parameters and Jones vector, we deduce a polarization evolving state and a polarization locked state in strongly coupled region and weakly coupled region, respectively.

\end{abstract}

\pacs{Valid PACS appear here}
\maketitle


\section{Introduction}

Microresonator platforms have gained considerable research interest in the last decade due to its numerous applications and scopes \cite{Brasch, Lee}. Optical frequency comb (FC) and its temporal manifestation in the form of cavity solitons (CSs) are the results of light-matter interaction in these resonator structures \cite{Hansson, Herr}. Mode-locked CS-based FC has paved the way towards precision spectroscopy \cite{Weichman}, wavelength division multiplexed optical communication \cite{Fulop}, optical clocks \cite{Papp}, ultrafast optical ranging \cite{Trocha}, optical switches \cite{Guo} and so on.
\par \noindent Optical resonator supports multiple mode families depending on the number of total internal reflection a light ray suffers within a single round trip. These are called resonant modes $(M)$ of the system, which can be estimated from the relation $2\pi R n_{\text{clad}}/\lambda_p \leq M \leq 2\pi R n_{\text{core}}/\lambda_p$, (more specifically can be defined as, $M= 2\pi R n_{eff}/\lambda_p$ ) where, $n$ is the refractive index of the material and $R$ is the radius of the cavity \cite{Lam}. A stable single mode-locked pulse dynamics can be interpreted as a double balance between Kerr nonlinearity, dispersion, the external CW pump, and the total loss in the dissipative systems \cite{Haldar, Roy}. Moreover, anomalous dispersion is desirable to obtain a bright soliton pulse and broad FC. Resonators with normal dispersion, in general, hinders the possibility of CS, where coupling between two normal dispersion modes modifies the local dispersion of one mode to the anomalous region and helps to produce a stable, bright pulse in the same resonator system \cite{Fujii, Liu}. Researchers have studied the dynamics of the resonator field in the presence of various mode coupling phenomena such as coupling between clockwise-anticlockwise modes \cite{Fujii2,Yoshiki}, orthogonally polarized modes \cite{Saha, Suzuki, Hansson2} or modes which encounters avoided mode crossing (AMC) \cite{Fujii, Liu, Aguanno, Lucas}.
\newline \noindent In the present work, our sole interest is to study the temporal and spectral dynamics of the resonator field in presence of mode coupling near the AMC region in a concentric microring resonator. When two modes of a resonator system have an identical resonant frequency, a new pair of hybrid modes may appear due to mode coupling between them \cite{Wiersig, Soltani}. We find out the intra-cavity field dynamics in two regions i.e. strongly coupled hybrid mode region (SCR) and weakly coupled hybrid mode region (WCR) based on the eigenfrequency separation of the two-hybrid modes \cite{Soltani}. We observe oscillations of peak power as well as transfer of energy between the hybrid modes in the temporal domain and periodic variation of FC width in the spectral domain. We also witness non-identical and identical phase variations of the mode field in SCR and WCR respectively. \textcolor{black}{To distinguish the pump wavelength region within which hybrid modes will exist, we calculate the overlap integral (OI) between the two hybrid mode fields. Depending on the OI values, we obtain the boundary region for strongly coupled and weakly coupled hybrid modes. Outside the boundary region for weakly coupled hybrid modes, we can safely say that hybrid modes do not exist as their resonance frequencies are too far to have sufficient amount of spatial overlap.} We further verified the field dynamics theoretically with the help of \textit{Rayleigh's dissipation function} (RDF) in the framework of variational principle \cite{RoyJLT,Saha2,Sahoo}. We also study the polarization state of the intra-cavity field in these two separate regions with Stokes parameters and Jones vector.
\section{Resonator Modelling}
\begin{figure}[tp]

	\includegraphics[trim={0.37\textwidth} 0.0in {0.1\textwidth} 0.0in,clip=true,width=0.5\textwidth,height=0.35\textwidth]{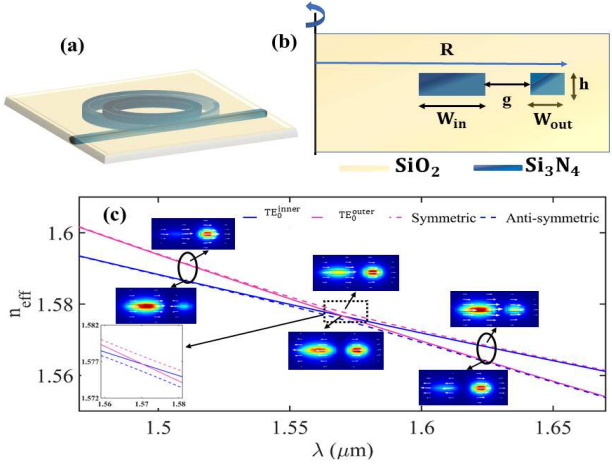} 
\caption{(a) Schematic of silicon nitride concentric microring resonator with silica cladding (b) 2D waveguide geometry of the same, W$_{\text{in}} = 2600$ nm, W$_{\text{out}} = 1085$ nm, g $= 900$ nm, h $= 300$ nm, R $= 50$ $\mu$m (c) Variation of effective refractive index (n$_{\text{eff}}$) profile with wavelength $(\lambda)$; n$_{\text{eff}}$ of individual fundamental modes of two resonators (TE$_0^{\text{inner}}$ and TE$_0^{\text{outer}}$) crosses around 1.56 $\mu$m (solid lines), mode coupling induced hybrid modes avoiding the crossing point (dotted lines)[shown in inset]. Symmetric mode makes a transition from outer to inner resonator around the AMC region as the wavelength increases and anti-symmetric mode undergoes the opposite transition.}
\label{fig.1}
\end{figure}
\noindent In this section, we model and characterize concentric microresonators exhibiting avoided mode crossing   \cite{Soltani, Kim}. The concentric microresonators is made of Si$_3$N$_4$ with SiO$_2$ cladding, shown in Fig.\ref{fig.1}(a). The structural parameters that we consider for this waveguide are: thickness (h) of the individual ring is $300$ nm, ring radius (R) is $50$ $\mu$m, widths (W$_{\text{in}}$ and W$_{\text{out}}$) are $2600$ nm and $1085$ nm for inner and outer rings, respectively, the gap ($g$) between the two rings is $900$ nm. The design geometry is shown in Fig.\ref{fig.1}(b). It is observed that in the absence of any mode-coupling, the effective refractive index ($n_{\text{eff}}$) of the fundamental transverse electric modes (TE$_0^{\text{inner}}$ and TE$_0^{\text{outer}}$ ) for two resonators crosses each other at a particular wavelength. In Fig.\ref{fig.1}(c), we plot $n_{\text{eff}}$ as a function of $\lambda$ where the cross-over takes place around $\lambda \approx 1.57$ $\mu$m (solid lines). This AMC position can be tuned with structural parameters like $g$ or W$_{\text{in}}$ and W$_{\text{out}}$. The hybrid mode, which moves from outer resonator to inner resonator with increasing wavelength, is known as symmetric mode and the mode which undergoes a transition from inner resonator to outer one, known as asymmetric mode, respectively \cite{Kim}. \textcolor{black}{We demonstrate the vectorial field profile of the electric field confined in the two hybrid modes (indicated by arrow lines). It is evident that, for the symmetric mode, the electric field orientation in two coupled resonators are in the same direction whereas, for anti-symmetric modes, the orientation of the field is in opposite direction and the field also reverses its direction in either side of the AMC point.} From the coupled-mode analysis for a loss-less resonator, one can readily write down the eigenfrequencies $(\omega^{\text{as,s}})$ of these two hybrid modes as \cite{Fujii}:

  \begin{figure}[tp]
\centering

\subfloat{
	\includegraphics[trim= {0.03\textwidth} {0.08\textwidth} 0.7in 0.8in,clip=true,width=0.48\textwidth,height=0.22\textwidth]{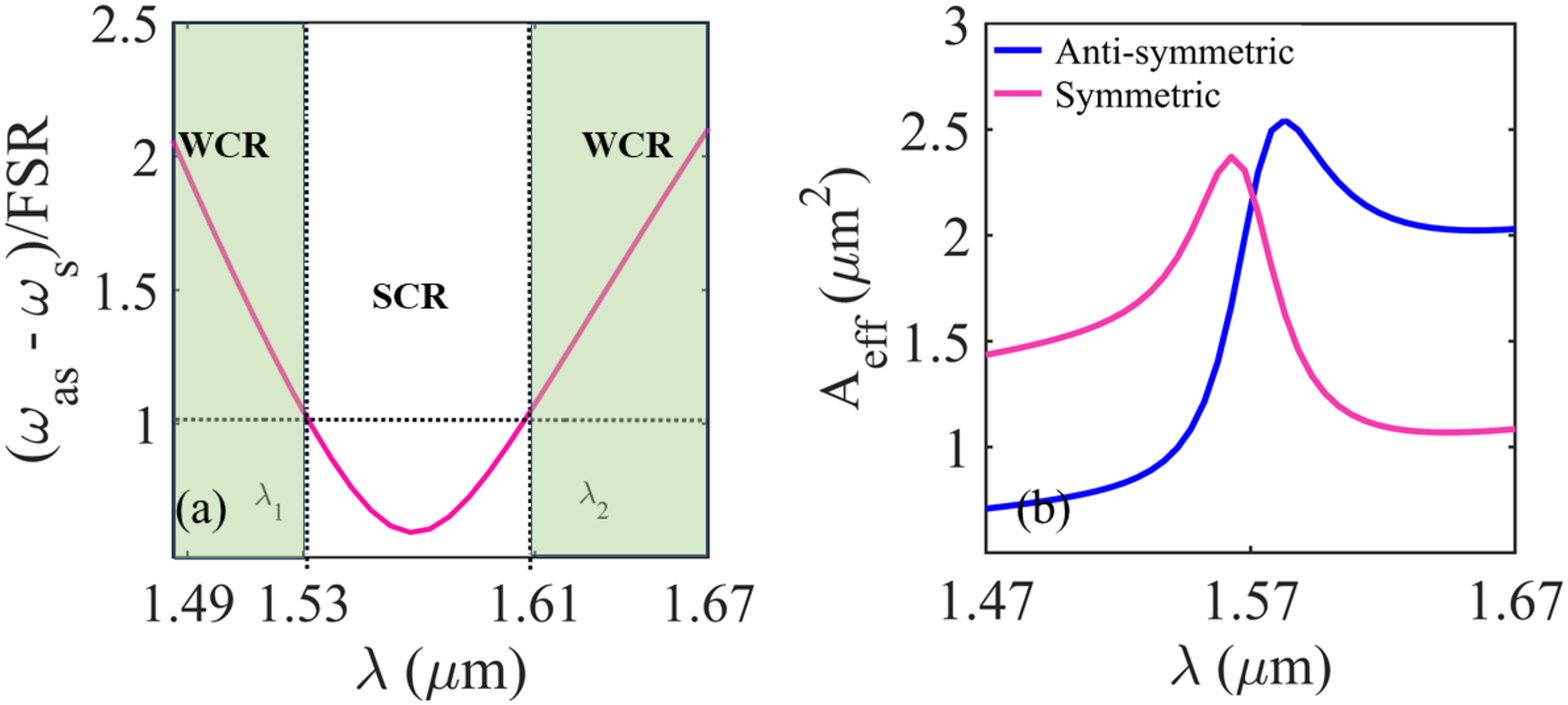}} 
	
 \subfloat{
	\includegraphics[trim= 0.5in 0.8in 0.3in {0.08\textwidth},width=0.48\textwidth,height=0.22\textwidth]{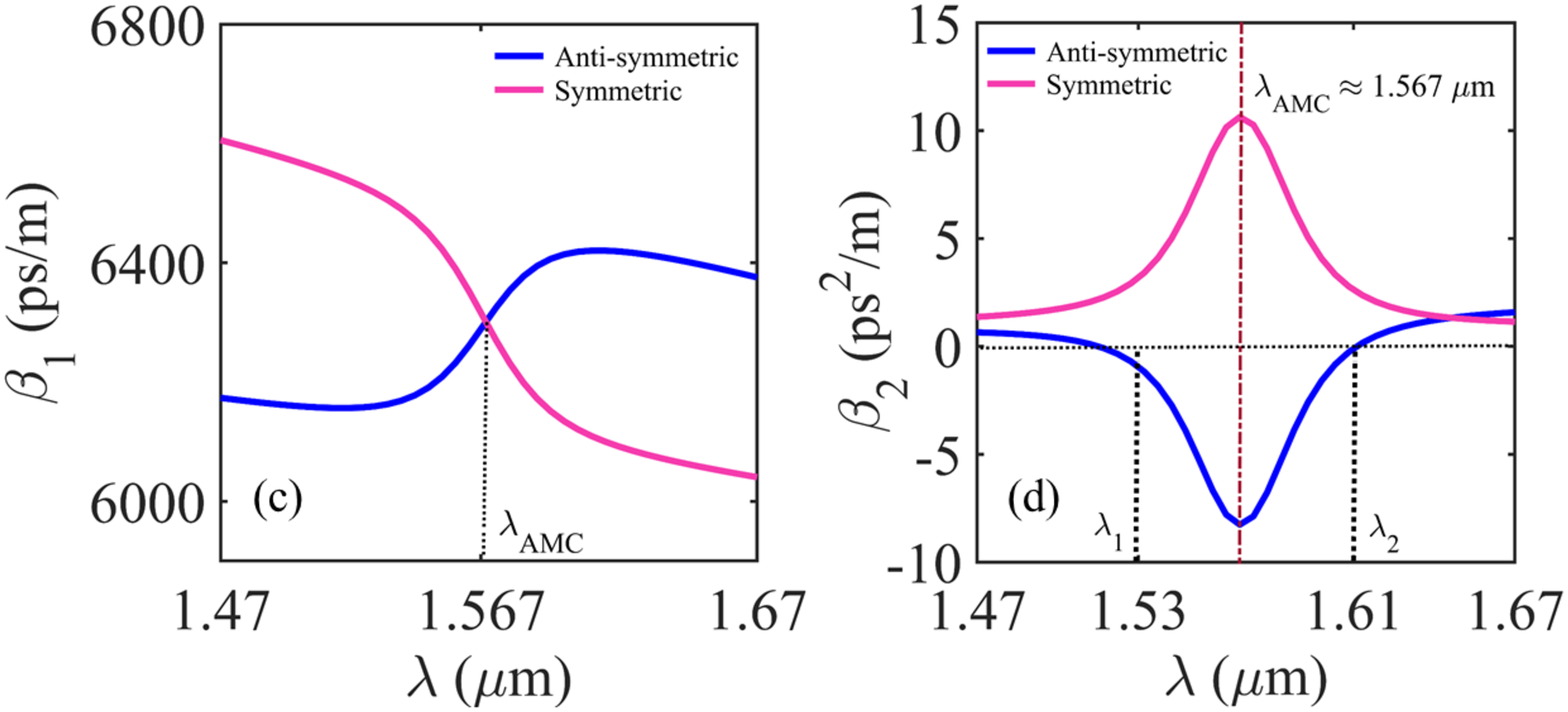}} 
 
\caption{(a) Normalized angular resonance frequency difference ($(\omega_{as}-\omega_s)/FSR_\text{outer}$) of two hybrid modes vs. the pump wavelength. $\lambda_1$ and $\lambda_2$ are the lower and upper cut-off values of the SCR, shaded regions represent weakly coupled hybrid mode region (WCR). (b) Effective mode area of two hybrid modes as function of pump wavelength. (c) Variation of inverse group velocities ($\beta_1$) of symmetric and anti-symmetric modes with pump wavelength. (d) Group velocity dispersion ($\beta_2$) of two hybrid modes, two dotted lines ($\lambda_1,\lambda_2$) marked for boundary of SCR.  }
\label{fig.2}
 
\end{figure}


\begin{equation}
    \omega^{\text{as,s}} = \frac{\omega_1+\omega_2}{2} \pm \sqrt{\frac{(\omega_1-\omega_2)^2}{4}+\frac{|\kappa|^2}{4}}.
    \label{eq.1}
\end{equation}

\par \noindent Exploiting the ($n_{\text{eff}}$-$\lambda$) plot obtained from the mode analysis in COMSOL, we calculate the eigenfrequencies of the modes both in the uncoupled $(\omega_1, \omega_2)$ and coupled conditions using the following relation $\omega = \frac{2\pi c}{n_{\text{eff}}\lambda}$. Thus, we readily obtain the mode coupling value $(\kappa)$ from Eq. (1). The resonant modes have slightly different coupling factor i.e., $\kappa_{1} \neq \kappa_{2}$ because the concentric ring waveguides are slightly different in size. We further use coupling coefficient $(\kappa)$ : $\kappa = \sqrt{\kappa_{1}\kappa_{2}}$ as defined in \cite{Okamoto}. In Fig.\ref{fig.2}(a), we demonstrate the \textit{resonance splitting} $ \Delta \omega_R=(\omega_{as}-\omega_s)$ of two hybrid modes which is normalized by the \textit{free spectral range} ($FSR_{\text{outer}}$) of the outer ring resonator. \textcolor{black}{We mark two regions for hybrid modes namely, (i) \textit {strongly coupled hybrid mode region (SCR)} (1.53 $\mu$m $\leq \lambda \leq$ 1.61 $\mu$m) and (ii) \textit{weakly coupled hybrid mode region (WCR)}(1.49 $\mu$m $\leq \lambda <$ 1.53 $\mu$m and 1.61 $ \mu$m $<\lambda \leq$ 1.67 $ \mu$m). In SCR, the two hybrid modes are correlated with the eigenmodes of each microring and can exist only when $\Delta \omega_R/FSR_{\text{outer}} \leq 1$. For WCR,  $1<\Delta \omega_R/FSR_{\text{outer}} <2$, indicating reduced coupling effect between the hybrid modes.} In Fig.\ref{fig.2}(b), we demonstrate the variation of effective mode area ($A_{\text{eff}}$) for both hybrid modes as a function of pump wavelength.  Note that, $A_{\text{eff}}$  attains a peak value near AMC wavelength. In Fig.\ref{fig.2}(c), we plot the variation of group velocity inverse for two modes, $\beta_{\text{1(as,s)}} = v_{\text{g(as,s)}}^{-1}$, where $v_{\text{g(as,s)}} $ represent the group velocities of corresponding hybrid modes. It is observed that at AMC wavelength, both the modes have nearly identical $\beta_1$ values. In Fig.\ref{fig.2}(d), we demonstrate the group velocity dispersion (GVD) profile for symmetric and anti-symmetric modes. It is interesting to note that under mode coupling condition, symmetric mode always remains in normal GVD regime. Conversely, in the SCR the anti-symmetric mode undergoes anomalous GVD regime. Vertical dotted lines mark the lower and upper cutoff values of the SCR. 
 \begin{figure}[bp]
\centering
	\includegraphics[clip,trim={0.01\textwidth} {0.01\textwidth} {0.01\textwidth} {0.01\textwidth},width=0.5\textwidth,height=0.28\textwidth]{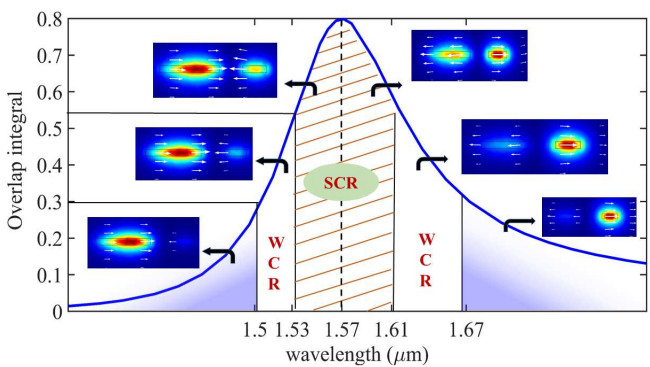}

\caption{(a) Overlap integral between two hybrid modes as a function of pump wavelength. Cut-off wavelength region for strongly coupled hybrid modes and weakly coupled hybrid modes are indicated, AMC wavelength is marked with dotted vertical line around 1.57 $ \mu$m. Gradient shaded region indicates nearly independent eigenmodes where the mode coupling effect is negligible. Vectorial anti-symmetric hybrid mode field distribution is shown at different region.}
\label{new_3}
\end{figure}
 \textcolor{black}{We calculate the overlap integral (OI) between the two hybrid mode fields exploiting the following expression \cite{Damask},
\begin{equation}
    OI = \frac{\iint_{S} \psi_{as}(x,y).\psi_{s}^*(x,y)\hspace{1 mm}dxdy}{\sqrt{\iint_{S} |\psi_{as}(x,y)|^2\hspace{1 mm}dxdy \iint_{S}  |\psi_{s}(x,y)|^2\hspace{1 mm}dxdy}}
    \label{1a.eq}
\end{equation}
The information of the spatial distribution of the hybrid modes ($\psi_{as,s}(x,y)$) are obtained from COMSOL. In Fig.(\ref{new_3}), we plot the OI values as a function of wavelength and identify the region where the hybrid modes exist (1.49 $ \mu$m $ \leq\lambda< 1.67$ $\mu$m). When OI $\geq$ 0.55, we found strong mode coupling effect between the hybrid modes. In the limit, 0.30 $\leq$ OI $<$ 0.55, intracavity dynamics follows the characteristics of WCR. The spatial overlap between the two hybrid modes are negligible for OI $<$ 0.30. In this region, the individual resonance frequencies are so far apart that the concept of hybrid modes nearly vanishes. Dotted vertical line around $\lambda \approx 1.57$ $ \mu$m signifies the position of AMC wavelength. We have also illustrate the vectorial field distribution for anti-symmetric hybrid mode in different region.}
 
\par \noindent In order to study the intra-cavity field dynamics, we obtain the fundamental mode $(\omega_{l0})$ family whose eigenfrequency is nearest to the pump frequency $(\omega_p)$. For two hybrid modes with different $n_{\text{eff}}$, the fundamental pump mode number $(l_{0\text{(as,s)}})$ and their eigenfrequency $(\omega_{l0})$ will be slightly different $(l_{0\text{(as,s)}} = \frac{n_{\text{eff(as,s)}}\omega_p R}{c})$ for a single $\omega_p$. One can expand the eigenfrequencies of the fundamental mode family $(l)$ using a Taylor series expansion to obtain for first \textit{N} elements \cite{Grelu}:
\begin{equation}
    \omega_l = \omega_{l0(\text{as,s})} + \displaystyle\sum_{k=1}^{N} \frac{\zeta_k}{k!}(l-l_{0(\text{as,s})})^k,
    \label{eq.2}
\end{equation}
 where $\omega_{l0\text{(as,s)}}$ is the eigenfrequency of the individual pump mode (nearest to the external pump frequency) and $\zeta_k = \frac{d^k\omega}{dl^k}|_{l=l_{0,\text{(as,s)}}}$. Now,  $\zeta_1 = \frac{d\omega}{dl}|_{l=l_{0,\text{(as,s)}}}=\Delta{\omega_{\text{FSR}}}$ is the FSR of the corresponding resonator supporting the hybrid mode at pump frequency. The resonator being dispersive, the eigenmodes are not exactly equidistant, $\zeta_2 = \frac{d^2\omega}{dl^2}|_{l=l_{0,\text{(as,s)}}}$ is the second order dispersion coefficient for the hybrid modes. $\zeta_2$ can be related to the well-known notation for GVD ($\beta_2$) as, $\zeta_2 = -\frac{c}{n_{\text{eff}}}\zeta_1^2\beta_2$ \cite{Grelu}.  We study field dynamics at different pump frequencies near the AMC region. At each time, resonant mode number changes with external pump frequency and correspondingly $\omega_{l0(\text{as,s})}$, $\Delta{\omega_{\text{FSR}}}$, dispersion coefficients are calculated from the Taylor expansion coefficients.

\section{Coupled Lugiato-Lefever Equation Analysis}

 \noindent The field dynamics inside a dissipative system like Kerr microresonator can be modelled by the well-known Lugiato-Lefever equation (LLE) \cite{Lugiato}, which incorporates the Kerr nonlinearity, dispersion, external pump source and the total loss of the system. Apart from all these parameters, this system being dissipative, the field dynamics is highly sensitive to the external parameters like frequency detuning, i.e., the difference between pump frequency and the nearest cavity mode resonant frequency ($\omega_p - \omega_{l0(\text{as,s})}$). Here, we use coupled LLE to find the dynamics of hybrid modes experiencing the AMC. The coupled LLE's are as follows:
 \begin{multline}
 \frac{\partial u_{as}}{\partial t} = \Bigl(-\frac{1}{2}\Delta\Gamma_{as} + i\sigma_{as}  + \textcolor{black}{ \Delta\zeta_{1}\frac{\partial }{\partial\theta}} +i\frac{\zeta_{2_{as}}}{2!}\frac{\partial^2 }{\partial \theta^2}\\ + i g_{0as}(|u_{as}|^2+\textcolor{black}{2|u_{s}|^2)}\Bigr)u_{as}+ i\frac{\kappa}{2} u_s + \frac{1}{2}\Delta\Gamma_{as}F_{as},
 \label{eq.3}
 \end{multline}
 
\begin{multline}
 \frac{\partial u_{s}}{\partial t} = \Bigl(-\frac{1}{2}\Delta\Gamma_{s} + i\sigma_{s}  - \textcolor{black}{ \Delta\zeta_{1}\frac{\partial }{\partial\theta}} +i\frac{\zeta_{2_{s}}}{2!}\frac{\partial^2 }{\partial \theta^2}\\ + i g_{0s}(|u_{s}|^2+\textcolor{black}{2|u_{as}|^2)}\Bigr)u_{s}+ i\frac{\kappa}{2} u_{as} + \frac{1}{2}\Delta\Gamma_{s}F_{s}.
 \label{eq.4}
 \end{multline}
 \noindent In the construction of coupled LLE, Eq.(\ref{eq.3})  and Eq.(\ref{eq.4}) ${u_{as}}$ and ${u_s}$ represent the field amplitudes of  anti-symmetric and symmetric modes, respectively. In this scenario, either one of the hybrid modes (anti-symmetric or symmetric) is pumped at any particular pump frequency depending on, which mode confines in the outer microring resonator. At the lower pump frequency, anti-symmetric mode is excited $(F_{as}\neq 0, F_{s} = 0)$  whereas for higher pump frequency, symmetric mode is excited $(F_{as} = 0, F_{s} \neq 0)$. \textcolor{black}{The coupling coefficient ($\kappa$) arises due to the degeneracy in eigenfrequency of two different resonator eigenmodes. Here, $\kappa$ is equivalent to the separation between the two hybrid modes, which varies with pumping wavelength near the AMC region suggesting the linear coupling effect between these two hybrid modes. Due to the sufficient amount of overlapping field between these modes near AMC, we also include nonlinear cross-phase modulation (XPM) effect (sixth term on the right hand side of coupled equations (Eqs.\ref{eq.3}-\ref{eq.4})). Though, for the choice of the external parameters in this case, we have seen the effect of XPM is negligible compare to the effect of linear coupling between the two modes (see,\textit{ Appendix \ref{appendix_XPM}}). We further neglect any coherent coupling terms. It is possible when light is confined within the resonator for a long time covering multiple round trips, such that the traverse path length is much greater than the corresponding beat length \cite{Agrawal}}. Here, $t$ is the slow time, $\theta$ is the azimuthal coordinate of the resonator, $\Delta\Gamma_{\text{as,s}}$ is the resonance linewidth of a mode which can be determined from the relation $\Delta\Gamma_{\text{as,s}}=\frac{\omega_{l0\text{(as,s)}}}{Q}$. The Q-factor of a cavity can be calculated as $Q=\frac{\omega_p T_\text{rt}}{l_\text{rt}}$ where \textcolor{black}{$T_\text{rt}=\frac{2\pi}{\zeta_1}\simeq 1.98$ ps} is the round trip time and $l_\text{rt}$ is the round trip loss. Round trip number can be estimated as, `total time confined/ round trip time'=$ t/T_{rt}$. The nonlinear coefficient $g_0$ is defined as, $g_{0\text{(as,s)}}=\frac{\hbar\omega_{l0\text{(as,s)}}^2n_2\zeta_{1\text{(as,s)}}}{2\pi n^2 A_{\text{\text{eff(as,s)}}}}$, $\hbar$, $n_2$, $n$, $A_{eff(as,s)}$ are the reduced Planck constant, Kerr coefficient, the linear refractive index of the medium and effective area of the mode, respectively. $\kappa$ is the mode coupling constant, $F_{\text{as,s}}$ is the external pump field defined as $(1/2)\Delta\Gamma_{\text{as,s}} F_{\text{as,s}}=\sqrt{\Delta\Gamma_{\text{as,s}}}\sqrt{\frac{P_\text{in}}{\hbar\omega_{l0,\text{(as,s)}}}}$, where $P_\text{in}$ is the input power. We observe linewidth of the hybrid modes are of the same order, i.e. $\frac{\Delta\Gamma_{as}}{\Delta\Gamma_{s}} \approx 1$. So, from now onward we designate $\Delta\Gamma_{as} \approx \Delta\Gamma_{s}=\Delta\Gamma$ $\approx 50 \text{ GHz}$. \textcolor{black}{$\sigma_{\text{as,s}} = \omega_p - \omega_{l0\text{(as,s)}}$} is the detuning between the pump and resonance frequencies of the resonator.  $\Delta\zeta_1 = \zeta_{1,as}-\zeta_{1,s}$ is the difference between cavity FSR for the two hybrid modes. The spectral variation of FSR for the hybrid modes follow the similar pattern of $\beta_1$ (Fig.\ref{fig.2}(c)), mentioned in \cite{Aguanno}. For numerical simulations, it is useful to  normalize the coupled LLE which takes the form,
 \begin{multline}
     \frac{\partial \psi_{as}}{\partial \tau} = \Bigl(-1  \textcolor{black}{- i\alpha_{as} +  D_1\frac{\partial}{\partial\theta}} +i D_{2as}\frac{\partial^2 }{\partial \theta^2} \\+ i(|\psi_{as}|^2+\textcolor{black}{2|\psi_{s}|^2)}\Bigr)\psi_{as}+ i\chi \psi_s + S_{as},
     \label{eq.5}
 \end{multline}
 
 \begin{multline}
     \frac{\partial \psi_{s}}{\partial \tau} = \Bigl(-1 \textcolor{black}{- i\alpha_s -  D_1\frac{\partial}{\partial\theta}} +i D_{2s}\frac{\partial^2 }{\partial \theta^2} \\+ i(|\psi_{s}|^2+\textcolor{black}{2|\psi_{as}|^2)}\Bigr)\psi_{s}+ i\chi \psi_{as} + S_{s},
     \label{eq.6}
 \end{multline}
 \noindent where the parameters are rescaled as : $\psi_{as}= \sqrt{\frac{2g_{0as}}{\Delta\Gamma}}u_{as}$, $\psi_{s} =\sqrt{\frac{2g_{0s}}{\Delta\Gamma}}u_{s}$, $\alpha_{\text{as,s}} = -\frac{2\sigma_{\text{as,s}}}{\Delta\Gamma}$, $D_{2\text{(as,s)}}= \frac{\zeta_{2\text{(as,s)}}}{\Delta\Gamma}$, $D_1 = \frac{2\Delta\zeta_1}{\Delta\Gamma}$, $\chi = \frac{\kappa}{\Delta\Gamma}$, $S_{\text{as,s}}= \sqrt{\frac{4g_{0\text{(as,s)}}P_\text{in}}{\Delta\Gamma^2\hbar\omega_p}}$ and  $\tau = \frac{\Delta\Gamma }{2}t$. For simplicity, in simulations, we have excluded all higher-order dispersion and nonlinearity terms. \textcolor{black}{We choose our initial wave-function of the form,
 \begin{equation}
     \psi_{as,s}= \eta_{as,s}\sech\Bigl(\eta_{as,s}\theta\Bigr),
     \label{eq.6new}
 \end{equation}
 where, $\eta_{as,s}=\sqrt{2|\alpha_{as,s}|}$ and $\alpha_{as,s}$ is the normalized detuning parameter for the two hybrid modes. For a particular pump wavelength, the difference between the eigenfrequencies of the hybrid modes are very small $(\omega_{l0\text{(as)}} \approx \omega_{l0\text{(s)}})$, such that the detuning value for both modes are nearly same. Here, the normalized detuning value $\alpha_{as,s}$ is considered as unity for both cases.} Further, we analyze two regions of strongly and weakly coupled hybrid modes on the basis of their eigenfrequency separation. We observe damped oscillatory motion of intra-cavity field due to mode coupling for a particular choice of detuning frequency. Phase and the polarization state of the cavity field evolve significantly in a different way in these two regions.
\subsection{Strongly Coupled Region}
\begin{figure}[tp]
\centering

\subfloat{
	\includegraphics[clip,trim={0.08\textwidth} 1.5in 0.8in 0.32in,clip=true,width=0.5\textwidth,height=0.23\textwidth]{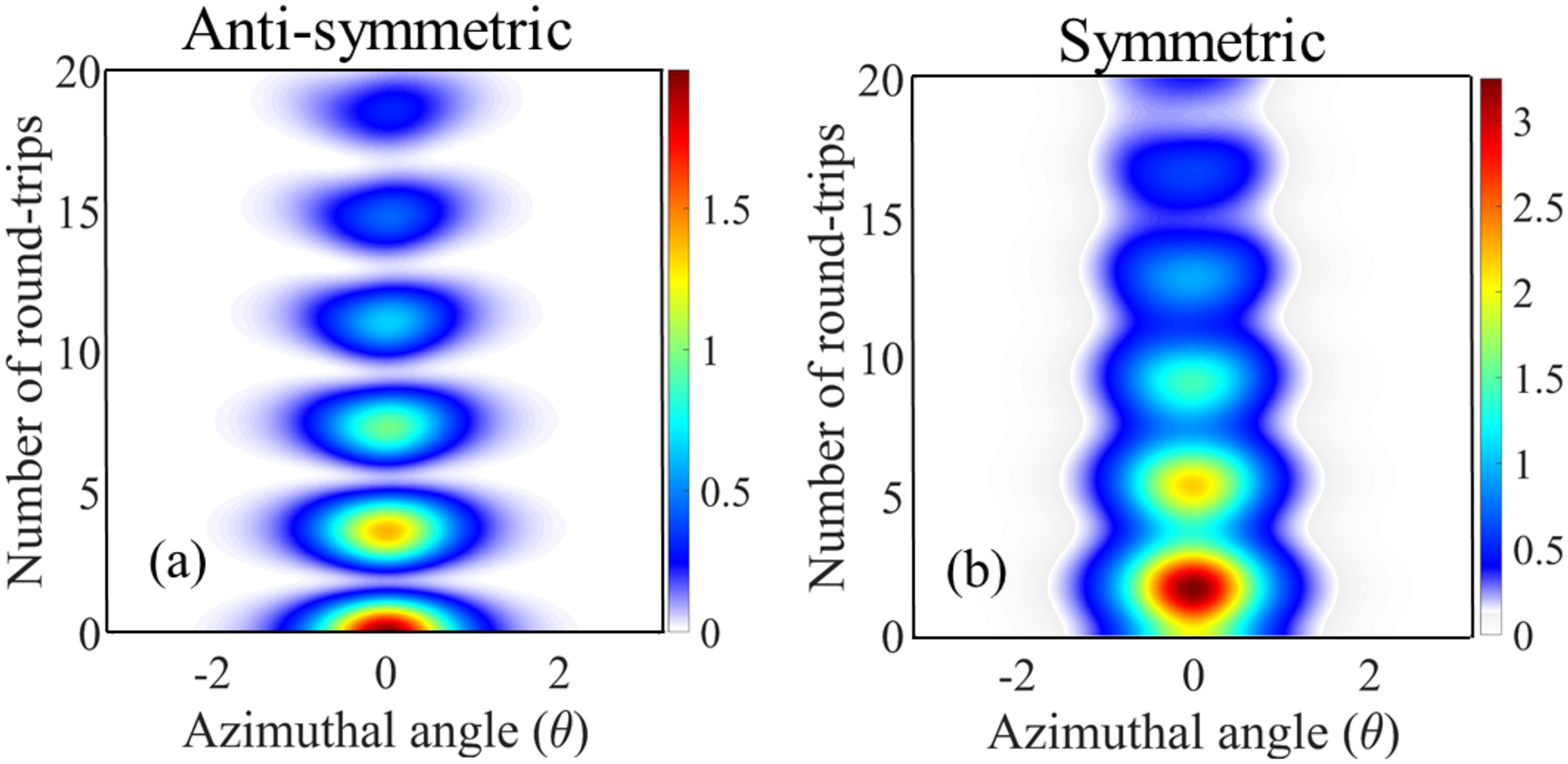}} 
 
 \subfloat{
	\includegraphics[clip,trim={0.03\textwidth} {0.2\textwidth} 0.6in {0.09\textwidth},width=0.5\textwidth,height=0.21\textwidth]{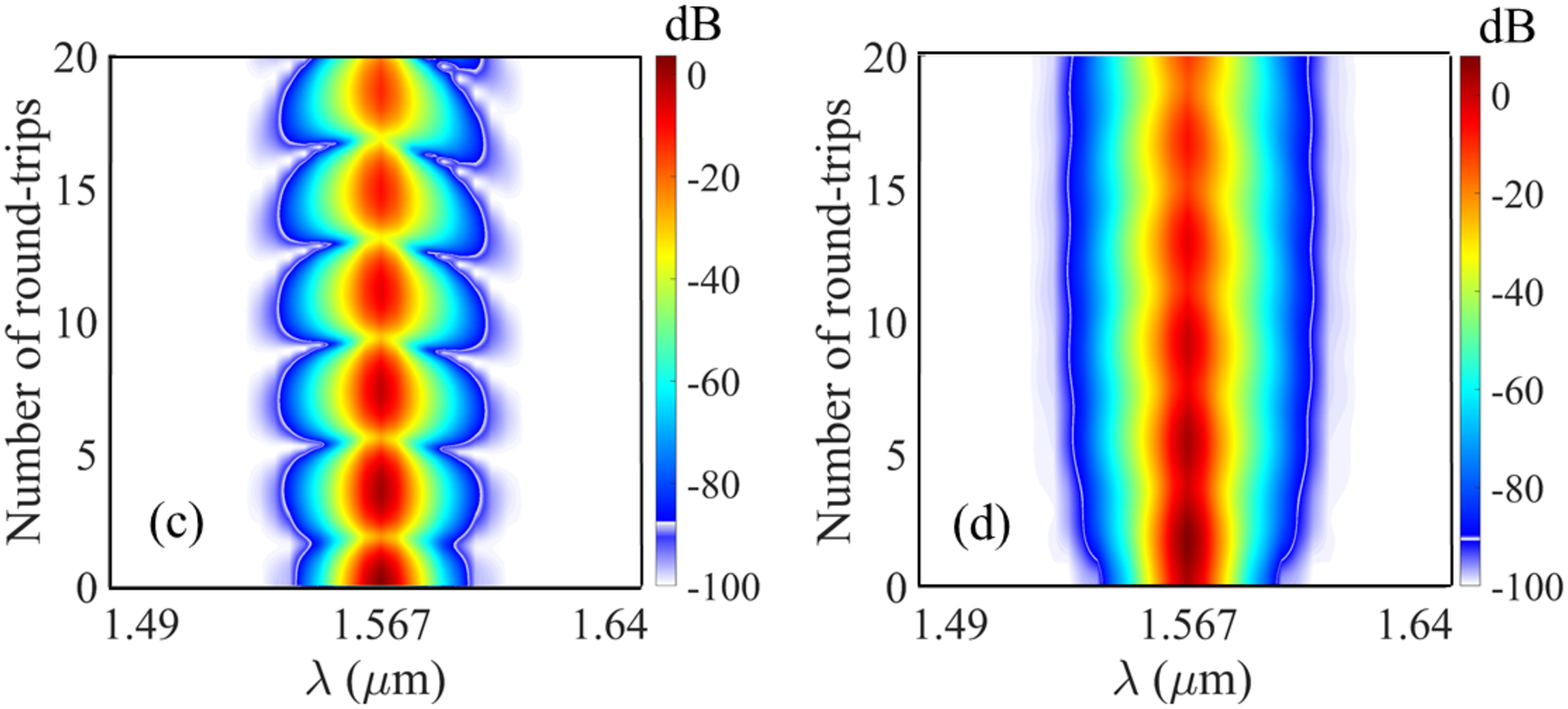}}

\caption{(a and b) \textcolor{black}{Temporal evolution of field and (c and d) evolution of frequency comb profile} of anti-symmetric and symmetric hybrid modes, respectively, at the AMC pumped near 1.567 $\mu$m. }
\label{fig.4_new}
\end{figure}
\begin{figure}[bp]
\centering
	\includegraphics[clip,trim={0.15\textwidth} {0.0\textwidth} {0.1\textwidth} {0.05\textwidth},width=0.5\textwidth,height=0.250\textwidth]{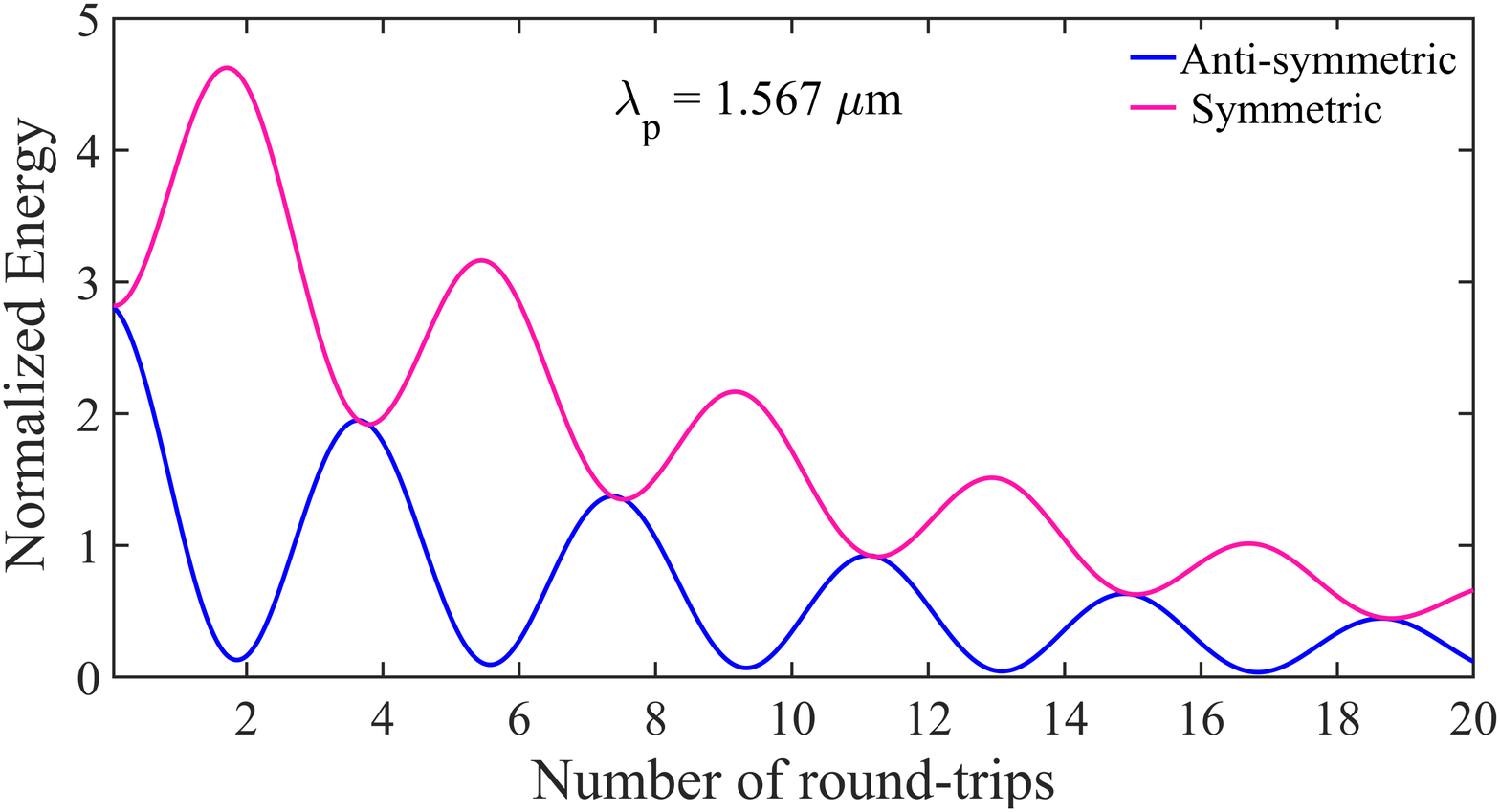}

\caption{Periodic transfer of energy between two hybrid modes with propagation. In both cases the initial energy is $2\sqrt{2}$.}
\label{fig.5_new}
\end{figure}
\noindent Strongly coupled hybrid modes exist within a frequency window where the difference between their eigenfrequencies i.e., $\Delta\omega_{R}=(\omega_{as}-\omega_s)$ is less than the FSR of the outer ring resonator, satisfying the condition $\frac{\Delta\omega_{R}}{FSR_\text{outer}}\leq 1$, shown in Fig.\ref{fig.2}(a). The main characteristics of SCR are; i) both hybrid modes show opposite dispersion i.e., the anti-symmetric mode falls in the anomalous dispersion while the symmetric mode experience normal dispersion. Further, the GVD attains its peak value ($'+ve'$ for symmetric, $'-ve'$ for anti-symmetric) at AMC (see Fig.\ref{fig.2}(d)), ii) GVM $(\delta = |1/\beta_{1,as}-1/\beta_{1,s}|)$ of two hybrid modes is minimum at AMC. In Fig.(\ref{fig.4_new}), we demonstrate the overall field dynamics at the AMC by numerically solving the coupled LLE. The parameters obtained from the structure at this pump wavelength are: $\zeta_{1,as}=3.1771$ THz, $\zeta_{2,as} =  1.714$ GHz, $\zeta_{1,s}= 3.1720$ THz, $\zeta_{2,s} =  -3.420$ GHz, $\kappa = 1.024$ THz. \textcolor{black}{We use cw pump power $P_\text{in} = 0.5$ W to the system.} The resonator eigenfrequency near the pump mode $(\omega_{l0\text{(as,s)}})$ is obtained from Eq.(\ref{eq.2}). \textcolor{black}{At the AMC wavelength, we obtain $\omega_{l0(as)} = 1.20315\times10^{15}$Hz, $\omega_{l0(s)} = 1.20247\times10^{15}$Hz.}  In Fig.\ref{fig.4_new}((a)-(b)), we depict the normalized peak power evolution for temporal field for both symmetric and anti-symmetric modes at the avoided crossing point.  The spectral counterpart showing the frequency comb width also periodically varies with propagation, shown in Fig.\ref{fig.4_new}((c)-(d)). We observe the peak power confined in each mode gradually decays with round trip time but with an oscillatory fashion. \textcolor{black}{ This periodic peak power evolution is not due to the breathing pulsation rather due to periodic energy exchange between modes. In order to confirm the periodic energy exchange we numerically calculate the total energy confined in each hybrid mode as, $E=\int_{-\infty}^{\infty} |\psi_{as/s}|^2\hspace{1 mm}d\theta$.  In Fig.(\ref{fig.5_new}), we plot the evolution of the total energy of the symmetric and anti-symmetric modes. Initially, the energy confined in both hybrid modes are equal, with propagation there is a periodic exchange of energy between them. In both the cases (symmetric and anti-symmetric) the overall energy decays with round trip exhibiting the transient nature of the mode in AMC region.} 
\begin{figure}[tp]
\centering
	\includegraphics[clip,trim={0.32\textwidth} {0.0\textwidth} 2.5in {0.01\textwidth},width=0.5\textwidth,height=0.40\textwidth]{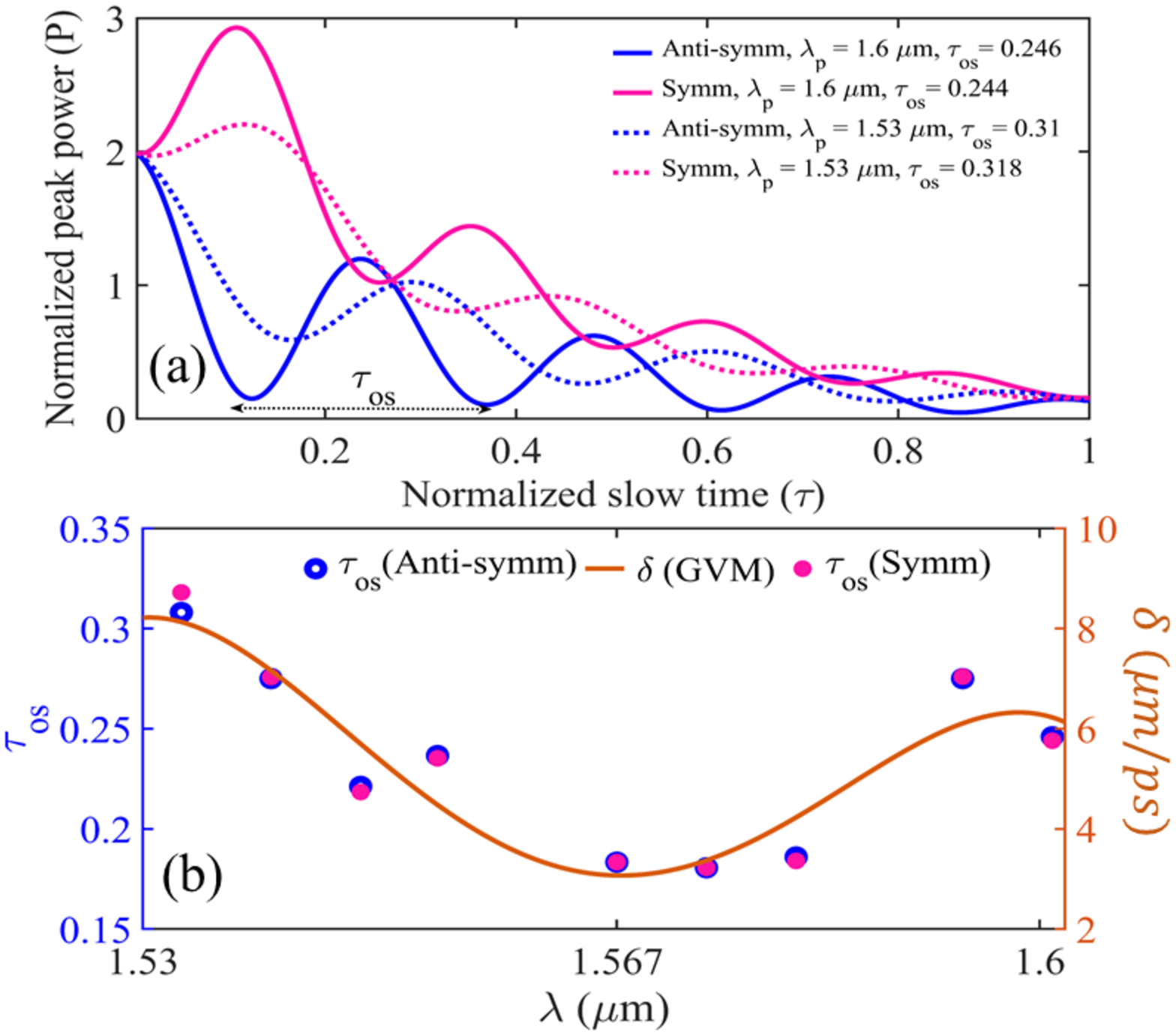}

\caption{(a) Power oscillations between symmetric and anti-symmetric modes at two edge pump wavelength ($\lambda = 1.53$ $\mu$m, $\lambda = 1.6$ $\mu$m) of strongly coupled region (SCR). (b) Variation of oscillation period $(\tau_{os})$ with pump wavelength within SCR for anti-symmetric mode (blue hollow circle) and symmetric mode (pink filled circle), solid curve shows the absolute GVM ($\delta$) variation between the hybrid modes with the pump wavelength.  }
\label{fig.6new}
\end{figure}
In Fig.\ref{fig.6new}(a), we numerically demonstrate the evolution of the peak power ($P$) of two modes with slow time ($\tau$) at two extreme wavelengths of SCR. We observe that these two modes' power remains correlated even at the edge wavelengths in either side of AMC, but oscillate with different oscillation periods. The oscillation period varies with the pump wavelength $\lambda_p$. In Fig.\ref{fig.6new}(b), we map the oscillation period $\tau_{os}$ for anti-symmetric (blue hollow circles) and symmetric (pink filled circles) modes field with operating wavelength and notice that $\tau_{os}$ attends its minima near the AMC wavelength ($\lambda_{AMC}$). We further find that the variation of  $\tau_{os}$ closely follows the nature of absolute GVM ($\delta = |\frac{1}{\beta_1,as}-\frac{1}{\beta_1,s}|$) between these two hybrid modes. Like  $\tau_{os}$, $\delta$ attends its minima at $\lambda_{AMC}$ and increases in either sides of $\lambda_{AMC}$. This correspondence leads us to the conclusion that GVM plays a critical role during the field oscillations.

\subsection{Weakly Coupled Region}
\begin{figure}[bp]
\centering

\subfloat{
	\includegraphics[clip,trim={0.06\textwidth} 1.3in 0.8in 0.6in,clip=true,width=0.5\textwidth,height=0.23\textwidth]{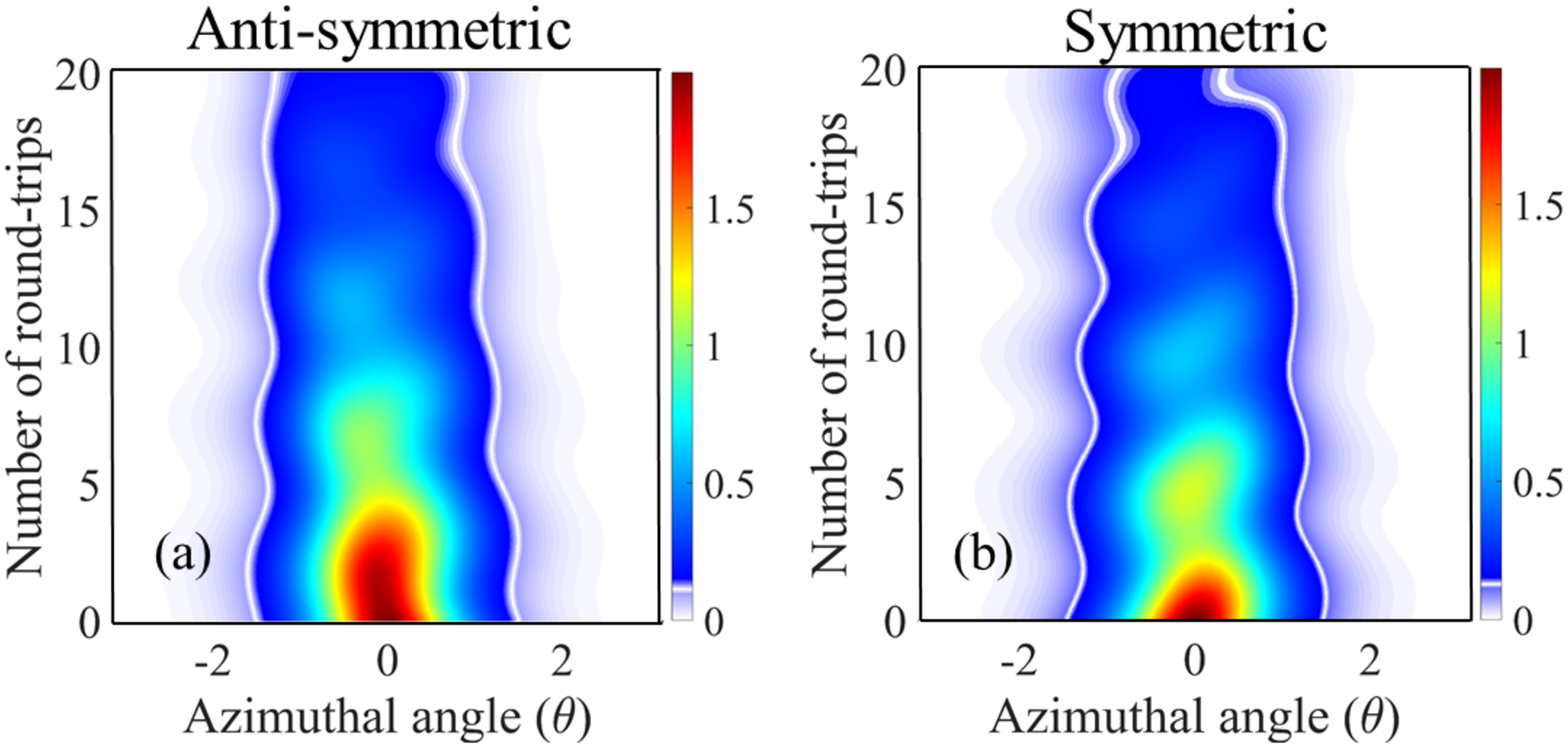}} 
 
 \subfloat{
	\includegraphics[clip,trim={0.03\textwidth} {0.2\textwidth} 0.6in {0.09\textwidth},width=0.5\textwidth,height=0.21\textwidth]{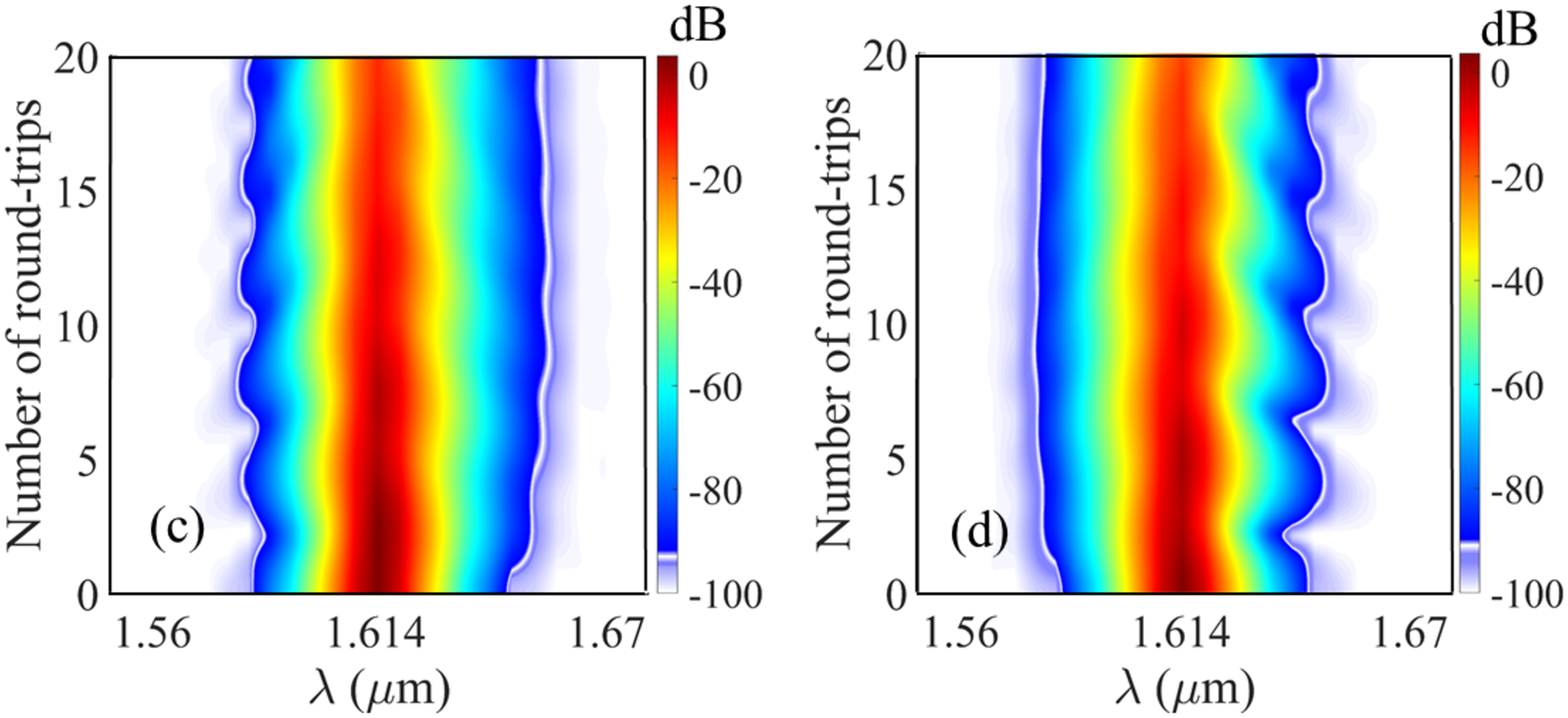}}
	
\caption{(a and b) Temporal and (c and d) spectral profile of anti-symmetric and symmetric hybrid modes, respectively, pumped at 1.614 $\mu$m in weakly coupled region.}
\label{fig7}
\end{figure}
\noindent In this subsection, we  discuss the basic characteristics of the hybrid modes when their coupling strength gradually reduces. This weakly coupled region lies for the pump wavelength \textcolor{black}{1.49 $\mu$m $\leq \lambda <$ 1.53 $\mu$m and 1.61 $ \mu$m $<\lambda \leq$ 1.67 $ \mu$m} as depicted by the shaded region in Fig.\ref{fig.2}(a). \textcolor{black}{We characterize the WCR which satisfies the two following conditions: i) where, $\Delta \omega_{R}$ satisfies the condition $1<\frac{\Delta \omega_{R}}{FSR_\text{outer}} <2$ and ii) overlap integral (OI) between the two hybrid modes lies within the range, $0.3\leq OI < 0.55$}.  In Fig.(\ref{fig7}), we demonstrate the temporal and spectral characteristics of hybrid modes when pumped in WCR $(\lambda_p = 1.614$ $\mu$m). The parameters used in the simulation  are as follows:  $\zeta_{1,as}= 3.1433$ THz, $\zeta_{2,as} =  1.7154$ GHz, $\zeta_{1,s}= 3.2456$ THz, $\zeta_{2,s} =  -3.2577$ GHz, $\kappa = 1.174$ THz, \textcolor{black}{cw pump power used here same as before (0.5W). Now, depending on the pump wavelength, whichever hybrid mode stays in the outer resonator gets directly pumped ($S_{\text{as/s}}\neq 0$) and the other mode is evanescently coupled to the mode present in the inner resonator ($S_{\text{as/s}} = 0$).} Here, We also obtain the eigenfrequencies $(\omega_{l0\text{(as,s)}})$ using Eq.(\ref{eq.2}). \textcolor{black}{At $\lambda_p = 1.614$ $\mu$m, we obtain $\omega_{l0(as)} = 1.167441\times10^{15}$Hz, $\omega_{l0(s)} = 1.167487\times10^{15}$Hz.} In Fig.\ref{fig7}(a-b) and Fig.\ref{fig7}(c-d), we represent the temporal and spectral characteristics of anti-symmetric and symmetric modes respectively for the aforementioned parameters. Unlike the situation at AMC (see Fig.(\ref{fig.4_new})), temporal field evolution is i) asymmetrical as the GVM increases, the field is either gradually ahead (anti-symmetric mode) or delayed (symmetric mode) from its initial position, ii) periodic exchange of energy between the hybrid modes gradually reduces and decays nearly in an identical way with propagation. We observe similar asymmetries in the spectral domain as well due to the increased GVM. Either left (anti-symmetric mode) or right (symmetric mode) wing width of the frequency comb oscillates with propagation. \textcolor{black}{In the other side of WCR, i.e. for 1.49 $\mu$m $\leq \lambda <$ 1.53 $\mu$m}, the only difference in the field characteristics from the one that are reported in Fig.(\ref{fig7}) is the nature of asymmetries in temporal and spectral profiles. The position of the asymmetries will be reversed due to the opposite GVM faced by two-hybrid modes in this side of AMC point.

 \section{Variational analysis}
 \noindent In this section, we introduce the Lagrangian perturbative approach based on variational principle \cite{Anderson, RoyJLT} to analyze the field dynamics semi-analytically.  The treatment requires a suitable ansatz whose parameters may evolve with slow time. Note, a major approximation of this treatment is the intactness of the ansatz function.
 We proceed with standard Kerr soliton ansatz having the mathematical form:
 \begin{equation}
     \psi_{as}=\eta_{as} \text{sech}(\eta_{as}\theta)e^{ i(\phi_{as}+ D_1 \theta)}
     \label{eq.7}
 \end{equation}
 \begin{equation}
     \psi_{s}=\eta_{s} \text{sech}(\eta_{s}\theta)e^{ i(\phi_{s}- D_1 \theta)},
     \label{eq.8}
 \end{equation}
 \noindent where the amplitudes $(\eta_{as}, \eta_{s})$ and the phases $(\phi_{as}, \phi_{s})$ of two CSs are allowed to vary over slow time, $\tau$. $D_1$ is proportional to the group-velocity mismatch between the two hybrid modes defined as, $D_1 \approx \delta/L_r$, where $L_r$ is the resonator length \textcolor{black}{and $\delta$ is the group-velocity mismatch between the two hybrid modes. As a standard procedure we define the Lagrangian in such a way that the following \textit{Euler-Lagrangian} (EL) equation retrieve the governing Eq.(\ref{eq.5}) and Eq.(\ref{eq.6}). Note, for mathematical simplicity, we ignore the negligible effect of nonlinear coupling terms in Lagrangian formalism.}
 \begin{equation}
\frac{d}{d\tau}\left(\frac{\partial L}{\partial \psi_{\tau}^*}\right)+\frac{d}{d \theta}\left(\frac{\partial L}{\partial \psi_{\theta}^*}\right)-\frac{\partial L}{\partial \psi^*}+\left(\frac{\partial R}{\partial \psi_{\tau}^*}\right) \\= 0,
\label{eq.9}
 \end{equation}
 \noindent  where $\textit{L}$ and $\textit{R}$ consist of:
 \begin{equation}
     L = L_{as} + L_{s} ,
     \label{eq.10}
 \end{equation}
 \begin{equation}
     R = R_{as} + R_{s}.
     \label{eq.11}
 \end{equation}
  \noindent  Here ($L_{as}$, $R_{as}$) and ($L_{s}$, $R_{s}$) represent the Lagrangian function and RDF of the anti-symmetric and symmetric modes, respectively.  Next, we reduce the  Lagrangian and RDF as, $ L_g=\int_{-\infty}^{\infty} L\hspace{1 mm}d\theta$ and $R_g=\int_{-\infty}^{\infty} R\hspace{1 mm}d\theta$ (details are given in \textit{Appendix \ref{appendix}}). Now we use the reduced form of $L_g$ and $R_g$ in EL equation:
\begin{equation}
\frac{d}{d\tau}\left(\frac{\partial L_g}{\partial \dot{p_j}}\right)-\frac{\partial L_g}{\partial p_{j}}+\left(\frac{\partial R_g}{\partial \dot{p_j}}\right) = 0,
\label{eq.12}
\end{equation}

\noindent where $p_j=\eta_{as},\eta_{s},\phi_{as},\phi_{s}$ and $\dot{p_{j}}= \frac{\partial\eta_{as}}{\partial\tau},\frac{\partial \eta_{s}}{\partial\tau},\frac{\partial\phi_{as}}{\partial\tau},\frac{\partial\phi_{s}}{\partial\tau}$, we obtain the coupled ODEs for two hybrid mode field parameters. The coupled differential equations governing the amplitude and phase of these fields are as follows:
 \begin{figure}[bp]
	\centering
	
	\includegraphics[clip,trim={0.06\textwidth} {0.1\textwidth} 1.2in {0.16\textwidth},width=0.5\textwidth,height=0.23\textwidth]{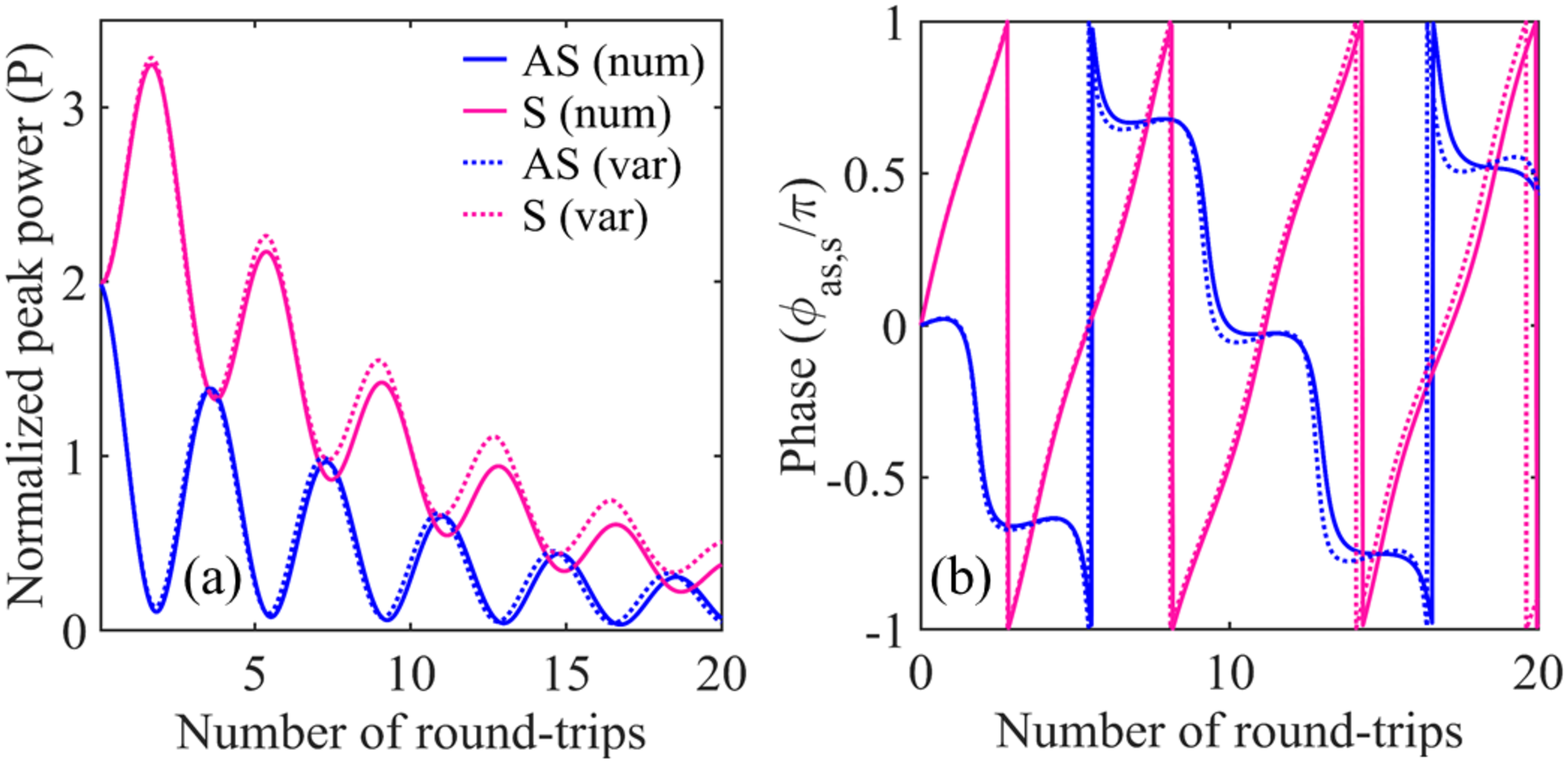}
	
	\caption{(a) Peak power and (b) phase variation for symmetric and anti-symmetric mode shown both numerically (solid lines) and with variational analysis (dotted lines) pumped at $\lambda=1.567$ $\mu$m in SCR. }
	\label{fig8}
\end{figure}
\begin{multline}
    \frac{\partial\eta_{as}}{\partial\tau}=\textcolor{black}{-2\eta_{as}} \underline{ -\chi\eta_{s}\Delta_{as}\sin\phi\csch\left(\Delta_{as}\right)}\\+ S_{as}\pi\cos\phi_{as}\sech\left(\frac{\Delta_{as}}{2}\right),
    \label{eq.13}
\end{multline}
\begin{multline}
    \frac{\partial\eta_{s}}{\partial\tau}=\textcolor{black}{-2\eta_{s}} \underline{ +\chi \eta_{as}\Delta_s\sin\phi\csch\left(\Delta_s\right)}\\+ S_{s}\pi\cos\phi_{s}\sech\left(\frac{\Delta_{s}}{2}\right),
    \label{eq.14}
\end{multline}    
\begin{multline}
    \frac{\partial\phi_{as}}{\partial\tau}=\textcolor{black}{-\frac{\Delta_{as}^2 S_{as}}{2D_1}\sin\phi_{as}\sech\left(\frac{\Delta_{as}}{2}\right)\tanh\left(\frac{\Delta_{as}}{2}\right)}\\ -\alpha_{as}+\textcolor{black}{D_1^2\left(1-D_{2as}\right)}+\eta_{as}^2\left(1-D_{2as}\right)\\-\chi\frac{\Delta_{as}\eta_{s}}{\eta_{as}}\cos\phi\csch\left(\Delta_{as}\right)\left(1-\Delta_{as}\coth\left(\Delta_{as}\right)\right),
    \label{eq.15}
\end{multline}
\begin{multline}
    \frac{\partial\phi_{s}}{\partial\tau}=\textcolor{black}{ -\frac{\Delta_s^2 S_{s}}{2D_1}\sin\phi_{s}\sech\left(\frac{\Delta_s}{2}\right)\tanh\left(\frac{\Delta_s}{2}\right)}\\ -\alpha_{s}+\textcolor{black}{D_1^2\left(1-D_{2s}\right)}+\eta_{s}^2\left(1-D_{2s}\right)\\-\chi\frac{\Delta_s\eta_{as}}{\eta_{s}}\cos\phi\csch\left(\Delta_s\right)\left(1- \Delta_s\coth\left(\Delta_s\right)\right),
    \label{eq.16}
\end{multline}
\begin{figure}[bp]
\centering

\includegraphics[clip,trim={0.06\textwidth} {0.1\textwidth} 1.2in {0.16\textwidth},width=0.5\textwidth,height=0.23\textwidth]{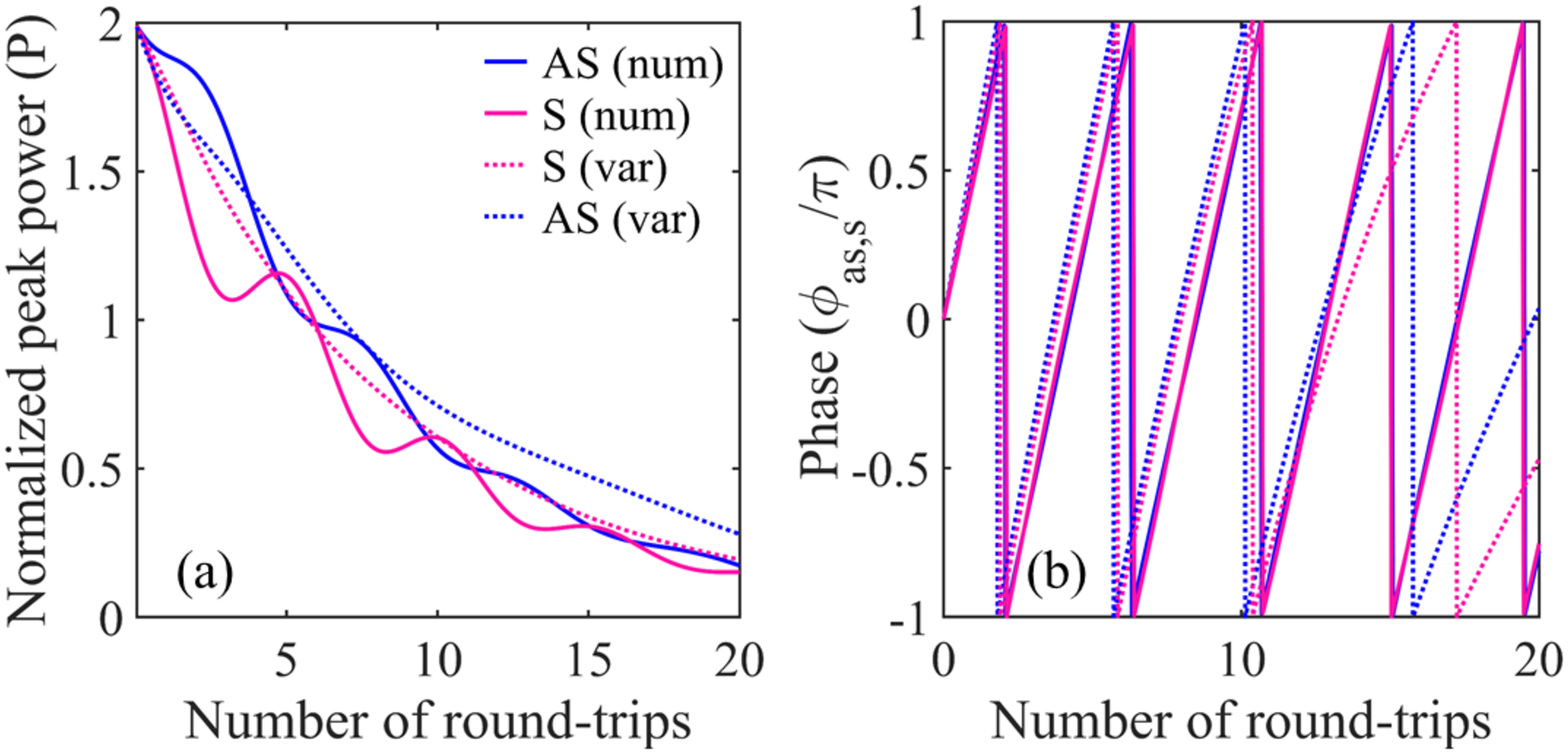}
	
\caption{(a) Peak power and (b) phase variation for symmetric and anti-symmetric mode shown both numerically (solid lines) and with variational analysis (dotted lines) pumped at $\lambda=1.614$ $\mu$m in WCR. }
\label{fig9}
\end{figure}
\begin{table*}[tp]
\caption{Polarized flux components evolution with normalized slow time in weakly and strongly coupled region.}
\centering
\begin{tabular}{|m{3.0cm}|m{4cm}|m{1.8cm}|m{1.8cm}| m{1.8cm}| m{1.8cm}| m{1.8cm}|}
\hline
Pump wavelength ($\lambda_p)$ & Measured Quantity & \multicolumn{5}{|c|}{Normalized slow time ($\tau$)}\\
\hline
\multicolumn{2}{|c|} {} & $\tau = 0$ & $\tau = 0.25$ & $\tau = 0.50$ & $\tau = 0.75$ & $\tau = 1.0$\\
\hline
\multirow{2}{8em}{1.567 $ \mu$m \\(SCR)}  & Degree of Linear polarization (DoLP) $\Bigl(\frac{\sqrt{S_1^2+S_2^2}}{S_0}\Bigr)$ & 1 & 0.80 & 0.54 & 0.92 & 0.87\\ & Degree of Circular polarization (DoCP) $\Bigl(\frac{|S_3|}{S_0}\Bigr)$ & 0 & 0.6 & 0.84 & 0.39 &0.48\\
\hline \hline
\multicolumn{7}{|c|} {} \\
\hline
\multirow{2}{8em}{1.614 $ \mu$m \\(WCR)}  & Degree of Linear polarization (DoLP) $\Bigl(\frac{\sqrt{S_1^2+S_2^2}}{S_0}\Bigr)$ & 1 & 0.999 & 0.998 & 0.998 & 0.999\\ & Degree of Circular polarization (DoCP) $\Bigl(\frac{|S_3|}{S_0}\Bigr)$ & 0 & 0.004 & 0.005 & 0.007 & 0.005\\
\hline 
\end{tabular}
\label{table1}
\end{table*}
\noindent where, \textcolor{black}{$\Delta_{as} = \frac{\pi D_1}{\eta_{as}}$, $\Delta_s = \frac{\pi D_1}{\eta_{s}}$,} $\phi = \phi_{s} -\phi_{as}$. These set of four coupled ordinary differential equations (Eq.\ref{eq.13}-Eq.\ref{eq.16})  capture the evolution dynamics of the amplitude and phase of the anti-symmetric $(\eta_{as},\phi_{as})$ and symmetric  $(\eta_{s},\phi_{s})$ modes respectively. In Fig.(\ref{fig8}), we have shown the field characteristics when pumped in SCR at AMC wavelength ($\lambda = 1.567 \mu$m). Fig.\ref{fig8}(a) shows the normalized peak power evolution for both hybrid modes. Simulation results (solid lines) are obtained by solving coupled LLE numerically. The results show how the power of the modes follows a damped oscillatory evolution. The symmetric and anti-symmetric modes accompany an inverse phase relationship dictating switching dynamics. We compare this simulation results with the theoretical prediction made through variational analysis and find an excellent agreement.  The underlined terms containing the coupling factor ($\chi$) in Eq.(\ref{eq.13}) and Eq.(\ref{eq.14}) responsible for the oscillatory pattern of the power evolution. The opposite sign of $\chi$ in two equations leads to the reverse patterns of oscillation between the two modes. The total amplitude $\eta (=\eta_{as}+\eta_{s})$ approximately follow an exponential decay as, $\eta \approx\eta_{0}e^{-\tau}$, when $\eta_{0}=\eta_{as}(0)+\eta_{s}(0)$. \textcolor{black}{Hence, for the initial wave-function given in Eq. \eqref{eq.6new}, we can have the total energy evolution $E_T(\tau)\approx4\sqrt{2}e^{-2\tau}$, where $E_T=E_{as}+E_{s}$. } In Fig.\ref{fig8}(b), we demonstrate the phase-evolution for symmetric and anti-symmetric modes. Variational results (dotted lines) support the numerical data (solid lines) to a good extent. Earlier, this non-identical phase variation near AMC has also been mentioned in \cite{Haus}. Note, we wrapped the phase value within the range $[-\pi,\pi]$. In Fig.\ref{fig9}(a,b), we plot the evolution of normalized peak power and phase with slow time, both numerically (solid lines) and with variational results (dotted lines), when the pump wavelength is in WCR ( $\lambda_p = 1.614$ $\mu$m). Numerical simulations reveal power oscillation to some extent within these hybrid modes; however, peak power follows exponential decay. In contrast to Fig.\ref{fig8}(a), here, the symmetric mode decays faster than the anti-symmetric mode initially, and after which, both of them decay nearly in a similar fashion. Variational results could not capture those power oscillations; instead, it results in exponential fall of amplitudes similar to the simulation results' envelope. 
\begin{figure*}[htbp]
\centering
\includegraphics[trim= 3.5in 0.2in 3.2in 0.3in,width=0.45\textwidth,height=0.38\textwidth]{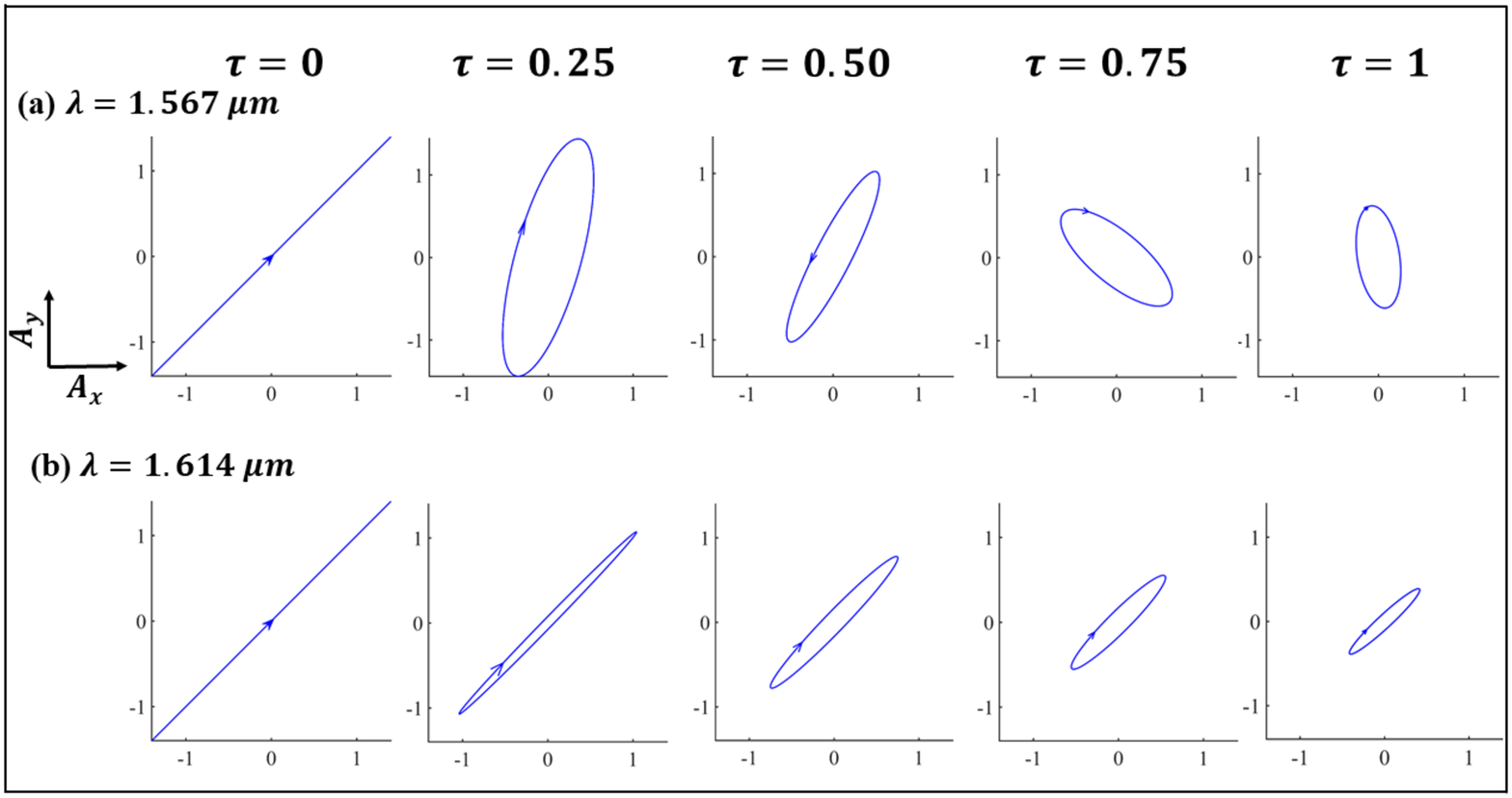}
\caption{Polarization ellipse evolution with slow time ($\tau$). (a) when the pump is in SCR at $\lambda = 1.567$ $\mu$m,  helicity reverses at $\tau = 0.50$, (b)  when the pump is in WCR at $\lambda = 1.614$ $\mu$m,  helicity maintains throughout the evolution.}
\label{fig10}
\end{figure*}
Another striking contrast between the two regions is the evolution of pulse phase $(\phi_{\text{as,s}})$, shown in Fig.\ref{fig8}(b) and Fig.\ref{fig9}(b). In SCR (see Fig.\ref{fig8}(b)), we found that the phase of the two-hybrid modes behaves non-identically, whereas, in WCR, phase evolution are nearly identical. We observe that numerical (solid lines) and variational results (dotted lines) differ quantitatively, but qualitatively both reflect the identical nature of hybrid modes' phase variation.


\section{Polarization state of intra-cavity field}
\noindent So far, we have seen the evolution of non-identical and identical phase variation for the two hybrid modes in SCR and WCR, respectively. Initially, we launch a linearly polarized light, which is coupled to these two hybrid modes in presence of mode coupling. In SCR, the relative phase difference between two field also changes with slow time. Thus, it is possible that the net field within the resonator created in the presence of the interference between the two hybrid mode field may experience a change from its initial polarization state. To find the polarization properties, we evaluate Stokes parameters \cite{Averlant} and corresponding Jones vector \cite{Chipman}. We numerically find out the S-parameters as: $S_0 = |\psi_{as}|^2+|\psi_{s}|^2$, $S_1 = |\psi_{as}|^2-|\psi_{s}|^2$, $S_2 = \psi_{as}^*\psi_{s}+\psi_{as}\psi_{s}^*$ and $S_3 = i(\psi_{as}^*\psi_{s}-\psi_{as}\psi_{s}^*)$. Physically, these parameters signify the amount of flux radiated in space along different orientations. $S_0$ signifies the total flux, $S_1$ and $S_2$ together define the total linear flux ($\sqrt{S_1^2+S_2^2}$) and $S_3$ defines how much amount of flux is circularly polarized. We further find out that the output flux is fully polarized in both SCR and WCR by satisfying the condition $S_0 = \sqrt{S_1^2+S_2^2+S_3^2}$. When a light beam is fully polarized, Jones vector helps to visualize the geometrical orientation of the electric field direction. Jones vector has four degrees of freedom, two indicating the amplitudes \textcolor{black}{$(A_{x},A_{y})$} and two correspond to their phase \textcolor{black}{$(\Phi_{x},\Phi_{y})$}. It can be represented in terms of S-parameters \cite{Saha}:  
\begin{equation}
\begin{matrix}
J = 
\end{matrix}
\begin{pmatrix}
\sqrt{\frac{S_0 +S_1}{2}}\\ \sqrt{\frac{S_0 -S_1}{2}} e^{i\tan ^{-1}\Bigl(\frac{S_3}{S_2}\Bigr)}
\end{pmatrix}
\label{eq.17}
\end{equation}
\noindent where \textcolor{black}{$A_{x}= \sqrt{\frac{S_0 +S_1}{2}}$, $A_{y}= \sqrt{\frac{S_0 -S_1}{2}}$ and $(\Phi_x-\Phi_y) = \tan ^{-1}\Bigl(\frac{S_3}{S_2}\Bigr)$}. Polarized flux components for two pump wavelengths, one at SCR (1.567 $ \mu$m), and another at WCR (1.614 $ \mu$m) are listed in Table \ref{table1}, which we have calculated from S-parameters, obtained through the coupled LLE at $\theta = 0$. The degree of linear polarization (DoLP) indicates how the electric field confines in one plane. Similarly, the degree of circular polarization (DoCP) characterizes how much fraction of the polarized flux is circularly polarized. We launch a linearly polarised light (DoLP = 1, DoCP = 0). Previously, in Fig.\ref{fig8}(a) and Fig.\ref{fig9}(a), we observe that the peak power gradually decays with slow time. Thus, the magnitude of ($S_0$), as well as other S-parameters, also decrease with slow time. Though the helicity of the circularly polarized light has not taken care during the evaluation of DoCP, this fact has been introduced while drawing the polarization ellipse with Jones vectors' help. Data of individual S-parameter evolution and corresponding Jones vector have been mentioned in \textit{Appendix} \ref{appendix2}. From Table \ref{table1}, we observe in SCR, the polarization state evolves with slow time, and the majority of the flux evolves periodically from linear to circular and again back to the linearly dominating flux component.
In contrast to that, in WCR, the initial linear polarization state remains nearly fixed, the circular polarisation component remains negligible compare to its linear component throughout the evolution. Thus in WCR, the net field remains in a polarization locked state. In Fig.(\ref{fig10}), we have shown the evolution of polarization ellipse with the normalized slow time drawn out of the Jones vector. While pumping at 1.567 $\mu$m in SCR, the helicity of circularly polarized flux transforms from right to left orientation near $\tau = 0.50$, then goes back to right circular orientation (see Fig.\ref{fig10}(a)), along with the change in dominating flux components (\textit{linear $\rightarrow$ circular $\rightarrow$ linear}). In Fig.\ref{fig10}(b), we observe in WCR ($\lambda = 1.614$ $\mu$m), the dominating flux components always remain linear, and the helicity is also maintained throughout.
\newline \noindent Recent studies have shown stable FC generation due to nonlinear \cite{Aguanno} mode coupling, as well as linear mode coupling induced multi-FSR FC generation \cite{Fujii} near AMC in an optical resonator. Choosing the external parameter like detuning frequency can lead to stable multi-FSR FC or unstable oscillatory single FSR FC generation. Though, various studies \cite{Fujii, Liu, Soltani, Kim} discuss the stable multi-FSR region, the dynamics in the oscillatory region has gained barely any attention. We find an experimental observation has been reported with power oscillations between two counter-propagating modes in a whispering gallery mode resonator \cite{Yoshiki}, which is close to our observed damped oscillation phenomena. 
\section{conclusion}
\noindent To conclude our work, we have reported the intra-cavity field dynamics in a concentric dual microring resonator system using coupled Lugiato-Lefever equation. We analyze mode coupling induced temporal and spectral field dynamics in the proximity of avoided mode crossing regions, namely in strongly coupled and weakly coupled hybrid mode regions, based on the their resonance frequency splitting. We further support the numerical findings with a semi-analytical variational treatment that justifies the observation of peak power oscillations or the energy transfer between the hybrid modes solely occurs due to linear mode coupling. The oscillation period varies with pump frequencies within the SCR, which seems to follow the nature of group velocity mismatch between the two hybrid modes. We observe that mode coupling induced phase-shift leads to a striking difference in polarization properties of the total intracavity field in these two regions. The non-identical phase evolution of two hybrid mode fields in SCR leads to a polarization evolving state for the total intra-cavity field. While pumping in WCR, the intra-cavity field nearly maintains its initial polarization state throughout the evolution. We also calculated the overlap integral factor between two hybrid modes to identify the region of their existence. We believe that our findings will enrich the understanding of mode coupling induced intra-cavity field dynamics in coupled micro-resonator system.

\section*{Acknowledgment}
\noindent The author M.S would like to thank  Ministry of Human Resource
Development, Government of India and IIT Kharagpur for funding to carry out her research work. Author SKV acknowledge the support received from project vide sanction no. DST/NM/NNETRA/2018(6)-IITKGP.
\appendix
\section{Comparison of linear and nonlinear mode coupling effect on pulse dynamics}
\begin{figure*}[tp]
\centering
\includegraphics[trim= 4in 3.2in 4in 0.2in,width=0.4\textwidth,height=0.3\textwidth]{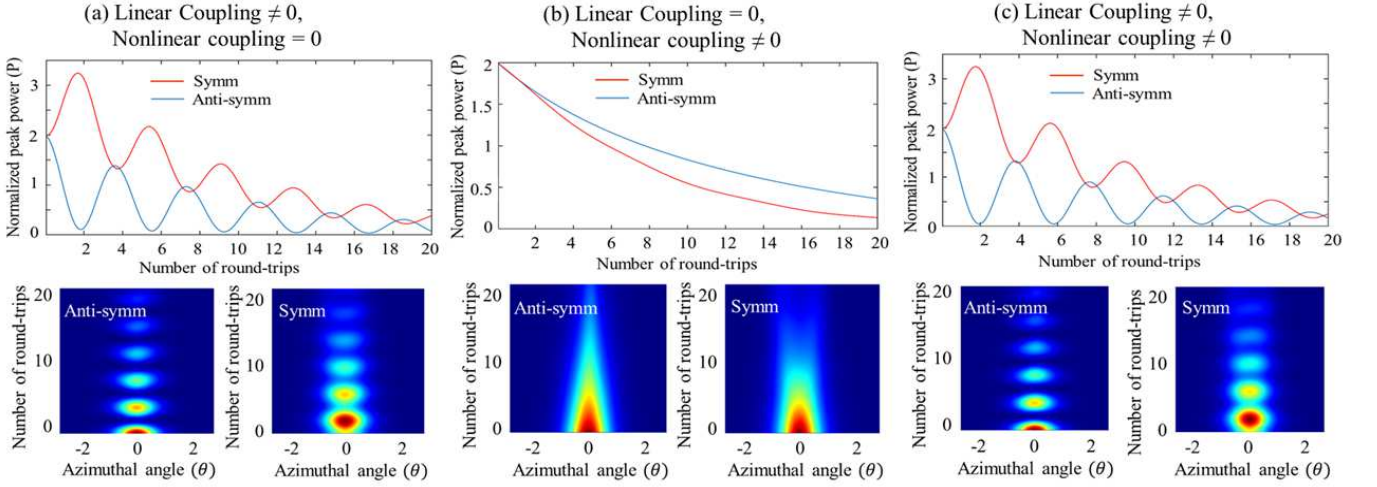}\vspace{1.5cm}
\caption{Comparison between the effect of linear and nonlinear (XPM) coupling during the evolution of the hybrid modes. Evolution of peak power (above) and total power evolution  (below) in two supermodes under (a) only linear coupling and (b) only nonlinear coupling (XPM) (c) both in presence of linear and nonlinear XPM coupling; pumped at $\lambda=1.567$ $\mu$m in SCR.  }
\label{fig11}
\end{figure*}

\noindent We have included linear and nonlinear cross-phase modulation (XPM) effect in the coupled LLE (Eq.\ref{eq.3}-Eq.\ref{eq.4}) to model hybrid mode evolution in the resonator. Here, we have used the XPM coefficient value to be `2' \cite{Aguanno}. In this section, we have shown the individual linear and nonlinear coupling effect in Fig.(\ref{fig11}). We can clearly see that in Fig.(\ref{fig11})(a) linear coupling introduces power oscillation among the supermodes, whereas, in
\label{appendix_XPM}
(b) XPM affects the coupling to the very beginning, after which pulse amplitude decays individually without any interaction. Total effect is shown in (c), which shows the dominant outcome of linear mode coupling between the two hybrid mode fields. As mentioned in \cite{Fujii2}, XPM is negligible for short pulse, and also because of choice of external parameters, we observe XPM induced phase-shift is quite negligible compare to the phase-shift due to linear mode coupling.

\section{Detailed form of Lagrangian and Rayleigh dissipation function}
\label{appendix}
\noindent The individual form of $L_{as}$,$L_{s}$ for the Lagrangian and $R_{as}$,$R_{s}$ for the RDF, are as follows:
\begin{multline}
L_{as}= \frac{i}{2}(\psi_{as}\psi_{as,\tau}^*-\psi_{as,\tau}\psi_{as}^*)\textcolor{black}{-\frac{i D_1}{2}(\psi_{as}\psi_{as,\theta}^*-\psi_{as,\theta}\psi_{as}^*)}\\+D_{2as}|\psi_{as,\theta}|^2-\frac{1}{2}|\psi_{as}|^4+\alpha_{as}|\psi_{as}|^2 -\frac{\chi}{2}(\psi_{s}\psi_{as}^*+\psi_{as}\psi_{s}^*),
\end{multline}
\begin{multline}
L_{s}= \frac{i}{2}(\psi_{s}\psi_{s,\tau}^*-\psi_{s,\tau}\psi_{s}^*)\textcolor{black}{+\frac{i D_1}{2}(\psi_{s}\psi_{s,\theta}^*-\psi_{s,\theta}\psi_{s}^*)}\\+D_{2s}|\psi_{s,\theta}|^2-\frac{1}{2}|\psi_{s}|^4+\alpha_{s}|\psi_{s}|^2-\frac{\chi}{2}(\psi_{as}\psi_{s}^*+\psi_{s}\psi_{as}^*),
\end{multline}
\begin{equation}
    R_{as} = -i S_{as}(\psi_{as,\tau}^*-\psi_{as,\tau})+i(\psi_{as}\psi_{as,\tau}^*-\psi_{as,\tau}\psi_{as}^*),
\end{equation}
\begin{equation}
    R_{s} = -i S_{s}(\psi_{s,\tau}^*-\psi_{s,\tau})+i(\psi_{s}\psi_{s,\tau}^*-\psi_{s,\tau}\psi_{s}^*),
\end{equation}
\noindent where, $\psi_{as,\tau/s,\tau}=\frac{\partial\psi_{\text{as,s}}}{\partial\tau}$ and $\psi_{as,\theta/s,\theta}=\frac{\partial\psi_{\text{as,s}}}{\partial\theta}$.\\\\
\noindent The following reduced forms of $L_g$ and $R_g$ are obtained after integrating the $\textit{L} (L_{as}+L_{s})$ and $\textit{R} ( R_{as}+R_s)$ over azimuthal angle ($\theta$), mentioned in the main text. In order to execute the integration, we have to assume the width of the  pulse in both modes remain same in the coupling term (containing $\chi$). Here, $\phi = \phi_{s} -\phi_{as}$.
\begin{multline}
    L_g = 2\eta_{as}\left(\frac{\partial\phi_{as}}{\partial\tau}-\textcolor{black}{D_1^2}+D_{2as}\left(D_1^2+\frac{\eta_{as}^2}{3}\right)-\frac{1}{3}\eta_{as}^2+\alpha_{as}\right)\\-2\chi\eta_{as}\Delta_s\cos\phi \csch \left(\Delta_{as}\right)\\+2\eta_{s}\left(\frac{\partial\phi_{s}}{\partial\tau}-\textcolor{black}{D_1^2}+D_{2s}\left(D_1^2+\frac{\eta_{s}^2}{3}\right)-\frac{1}{3}\eta_{s}^2+\alpha_{s}\right)\\-2\chi\eta_{s}\Delta_{as}\cos\phi \csch \left(\Delta_s\right),
\end{multline}
\vspace{1mm}
\begin{multline}
    R_g = \Bigl(4\eta_{as}-2S_{as}\pi\cos\phi_{as}\sech\left(\Delta_{as}\right)\Bigr)\frac{\partial\phi_{as}}{\partial\tau}\\-\left(\frac{\pi^2D_1S_{as} \sin\phi_{as}}{\eta_{as}^2}\sech\left(\Delta_{as}\right)\tanh\left(\Delta_{as}\right)\right)\frac{\partial\eta_{as}}{\partial\tau}\\+\Bigl(4\eta_{s}-2S_{s}\pi\cos\phi_{s}\sech\left(\Delta_{s}\right)\Bigr)\frac{\partial\phi_{s}}{\partial\tau}\\-\left(\frac{\pi^2D_1S_{s} \sin\phi_{s}}{\eta_{s}^2}\sech\left(\Delta_{s}\right)\tanh\left(\Delta_{s}\right)\right)\frac{\partial\eta_{s}}{\partial\tau},
\end{multline}
\section{Stokes parameters and Jones vector for pump wavelength in SCR and WCR}
\label{appendix2}
\noindent In this section, we have mentioned the Stokes parameters ($S_0$,$S_1$,$S_2$,$S_3$) and Jones vector ($J$) for two pump wavelengths, one in SCR $(\lambda =1.567$ $\mu$m), another in WCR $(\lambda =1.614$ $\mu$m). Stokes parameters are obtained 
\begin{table*}[tp]
\caption{Stokes parameters and Jones vector evolution with normalized slow time in strongly and weakly coupled region }
\centering
\begin{tabular}{|m{1.7cm}|m{1.4cm}|m{2.7cm}|m{2.7cm}| m{2.7cm}| m{2.7cm}| m{2.7cm}|}
\hline
Pump wavelength ($\lambda_p)$ & Measured Quantity & \multicolumn{5}{|c|}{Normalized slow time ($\tau$)}\\
\hline
\multicolumn{2}{|c|} {} & $\tau = 0$ & $\tau = 0.25$ & $\tau = 0.50$ & $\tau = 0.75$ & $\tau = 1.0$\\
\hline
\multirow{5}{8em}{1.567 $ \mu$m \\(SCR)}  & $S_0$ & 4 & 2.356 & 1.340 & 0.767 & 0.446\\ &$S_1$ & 0 & -1.781 & -0.754 & 0.086 & -0.310\\ &$S_2$ & 4 & -0.246 & 0.152 & 0.746 & -0.091\\ &$S_3$ & 0 & 1.523 & -1.098 & 0.156 & 0.307 \\ \cline{2-7} & J & 
$$
\begin{pmatrix}
1.414\\ 1.414e^{0.0i}
\end{pmatrix}
$$ & $$
\begin{pmatrix}
0.536\\ 1.438e^{-80.83i}
\end{pmatrix}
$$ & $$
\begin{pmatrix}
0.541\\ 1.023e^{-82.13i}
\end{pmatrix}
$$ & $$
\begin{pmatrix}
0.653\\ 0.583e^{11.84i}
\end{pmatrix}
$$ & $$
\begin{pmatrix}
0.260\\ 0.6148e^{-73.54i}
\end{pmatrix}
$$ \\  
\hline \hline
\multicolumn{7}{|c|} {} \\
\hline
\multirow{5}{8em}{1.614 $ \mu$m \\(WCR)}  & $S_0$ & 4 & 2.235 & 1.171 & 0.614  & 0.326\\ &$S_1$ & 0 & -0.066 & -0.039 & 0.001 & 0.022\\ &$S_2$ & 4 & 2.233 & 1.169 & 0.613 & 0.325\\ &$S_3$ & 0 & 0.028 & 0.025 & 0.02 & 0.013\\  \cline{2-7} & J & 
$$
\begin{pmatrix}
1.414\\ 1.414e^{0.0i}
\end{pmatrix}
$$ & $$
\begin{pmatrix}
1.041\\ 1.072e^{0.218i}
\end{pmatrix}
$$ & $$
\begin{pmatrix}
0.752\\ 0.779e^{.318i}
\end{pmatrix}
$$ & $$
\begin{pmatrix}
0.554\\ 0.553e^{.28i}
\end{pmatrix}
$$ & $$
\begin{pmatrix}
0.417\\ 0.389e^{.263i}
\end{pmatrix}
$$ \\
\hline 
\end{tabular}
\label{table2}
\end{table*}
from coupled LLE (Eq.\ref{eq.5}-Eq.\ref{eq.6}) at $\theta=0$, and the corresponding Jones vector ($J$) are obtained with the help of Eq.(\ref{eq.17}). We have shown the evolution of these two quantities with slow time in Table. \ref{table2}.

 \end{document}